\documentclass[aps,prl,twocolumn,superscriptaddress,appendix,showpacs,floatfix,nobibnotes]{revtex4}
\usepackage{epsfig}
\usepackage{epstopdf}

\usepackage{graphicx}
\usepackage{longtable}
\usepackage{CJK}
\usepackage{color}
\usepackage{mathptmx, courier, pifont}
\usepackage[scaled=0.92]{helvet}
\usepackage[T1]{fontenc}
\usepackage{textcomp}

\begin{document}

\begin{CJK*}{GB}{}

\title{The IBM Description of the B(E2) Anomaly: Dynamical Triaxiality and Configuration Mixing}

\author{Wei Teng}
\affiliation{Department of Physics, Liaoning Normal University,
Dalian 116029, P. R. China}

\author{Sheng Nan Wang}
\affiliation{Department of Physics, Liaoning Normal University,
Dalian 116029, P. R. China}

\author{Xian Zhi Zhao}
\affiliation{Department of Physics, Liaoning Normal University,
Dalian 116029, P. R. China}

\author{Yu Zhang }\email{dlzhangyu_physics@163.com}
\affiliation{Department of Physics, Liaoning Normal University,
Dalian 116029, P. R. China}

\begin{abstract}
A theoretical investigation has been carried out to examine the effects of dynamical triaxility and configuration mixing on the $B(E2)$ anomaly, characterized
by $B_{4/2}=B(E2;4_1^+\rightarrow2_1^+)/B(E2;2_1^+\rightarrow0_1^+)<1.0$ and $R_{4/2}=E(4_1^+)/E(2_1^+)\geq2.0$, within the framework of the interacting boson model (IBM). The results indicate that the effective $\gamma$ deformation may undergo substantial changes with increasing angular momentum in the $B(E2)$ anomaly system governed by an IBM Hamiltonian involving rotor-like terms. Further calculations reveal that mixing between normal and intruder states may facilitate the emergence of $B(E2)$ anomaly, thereby providing additional theoretical insights into this unusual phenomenon.
\end{abstract}

\maketitle

\end{CJK*}

\begin{center}
\vskip.2cm\textbf{I. Introduction}
\end{center}\vskip.2cm

The emergence of collective features represents one of the most remarkable characteristics of complex nuclear many-body systems.
The associated collective modes in even-even nuclei~\cite{Bohrbook} are expected to yield $B_{4/2}\equiv B(E2;4_1^+\rightarrow2_1^+)/B(E2;2_1^+\rightarrow0_1^+)>1.0$ along with $R_{4/2}=E(4_1^+)/E(2_1^+)\geq2$, which align well with various theoretical calculations and extensive experimental observations. However, recent measurements on some neutron-deficient nuclei near $N_\mathrm{n}=90$ and the proton dripline~\cite{Grahn2016,Saygi2017,Cederwall2018,Goasduff2019,Zhang2021,Zanon2025} reveal an anomalous collective phenomenon characterized by $R_{4/2}>2.0$ and $B_{4/2}\ll1.0$. To examine whether this exotic phenomenon, termed $B(E2)$ anomaly, can be understood from a collective model perspective, a series of theoretical analyses~\cite{Zhang2022,Wang2020,Zhang2024,Pan2024,Teng2025,Zhang2025} have recently been conducted within the framework of the interacting boson model (IBM)~\cite{Iachellobook}. The results suggest that strong band-mixing may act as a plausible mechanism underlying the $B(E2)$ anomaly observed in experiments. In these studies, the terms $LQL\equiv(\hat{L}\times\hat{Q}\times\hat{L})^{(0)}$ and $LQQL\equiv(\hat{L}\times\hat{Q})^{(1)}\cdot(\hat{L}\times\hat{Q})^{(1)}$ are introduced to incorporates the effects of band-mixing on $B(E2)$ transitions. Their connection to rotor modes in the IBM was recently highlighted in a more general context~\cite{Zhang2024,Zhang2025}. Indeed, the rotor-like terms and their roles in describing nuclear spectra have been well addressed within both the shell model method~\cite{Draayer1983,Draayer1989,Leschber1987,Castanos1988} and the IBM calculations~\cite{Heyde1984,Berghe1985,Vanthournout1988,Vanthournout1990,Smirnov2000,Zhang2014,Teng2024}. Only recently have the associated band-mixing effects been recognized as a means to model the observed $B(E2)$ anomaly features~\cite{Zhang2022,Wang2020,Zhang2024,Pan2024,Teng2025}, although the monotonically decreasing behavior of the $B(E2;L_1^+\rightarrow(L-2)_1^+)$ transitions with $B_{4/2}<1.0$ was early known~\cite{Zhang2014} to be producible from a finite-$N$ triaxial rotor mode. Typically, band mixing is associated with axially asymmetric systems~\cite{Wood2004,Allmond2008}. The resulting $B(E2)$ anomaly feature can naturally arise in a boson system with either intrinsic or dynamical $\gamma$ deformations, as analyzed in \cite{Zhang2025}. In particular, the latter case can manifest even within an axially deformed mean-field potential, provided there is considerable $\gamma$ softness. These findings are closely related to several mean-field calculations for relevant nuclei~\cite{Goasduff2019,Zanon2025,Guzman2010}, which show that different mean-field approaches can lead to distinct mean-field pictures for a $B(E2)$ anomaly nucleus. In addition, prior studies~\cite{Dracoulis1994,King1998,Harder1997,Morales2008,Ramos2009,Ramos2011,Ramos2014} have emphasized the importance of incorporating intruder configurations to accurately describe the spectral properties of Pt isotopes, where the $B(E2)$ anomaly feature has been recently observed in $^{172}$Pt~\cite{Cederwall2018}. In the IBM, intruder configurations induced by cross-shell excitation can be modeled through the configuration mixing scheme~\cite{Duval1982}, which is, however, not included in the previous IBM analysis of the $B(E2)$ anomaly.

In this work, we preform a theoretical analysis of the $B(E2)$ anomaly phenomena within the phenomenological version of the IBM~\cite{Iachellobook}, which does not distinguish between protons and neutrons. Instead of providing a new fit for the experimental data, the present analysis focuses on two aspects: how a rotor-like (band-mixing) term can influence dynamical triaxiality (effective $\gamma$ deformation)~\cite{Zhang2025}, and whether configuration-mixing can contribute to the occurrence of the $B(E2)$ anomaly.

\begin{center}
\vskip.2cm\textbf{II. Model Hamiltonian and Effective $\gamma$ Deformation}
\end{center}\vskip.2cm

In the IBM~\cite{Iachellobook}, the building blocks are two types of boson (operators): the monopole $s$ boson with $J^\pi=0^+$ and the
quadrupole $d$ boson with $J^\pi=2^+$. All physical operators, including the Hamiltonian, are constructed utilizing the creation and annihilation operators of these bosons.
According to the previous analyses~\cite{Zhang2025}, the $B(E2)$ anomaly induced by band-mixing can arise in an IBM system with either intrinsic or dynamic (effective) $\gamma$ deformations~\cite{Castanos1984,Casten1984,Elliott1986,Vogel1996}.
Apart from quantitative differences arising from variations in parameters or Hamiltonian forms, all models applied to $B(E2)$ anomaly~\cite{Zhang2022,Wang2020,Zhang2024,Pan2024,Teng2025,Zhang2025}
generally predict the presence of low-energies non-yrast states. This feature is typically regarded as a theoretical indicator of triaxiality~\cite{Bohrbook}.
For intrinsic axially asymmetric deformation, specific high-order terms must be introduced into the Hamiltonian to generate a $\bar{\gamma}\neq0$ mean-field picture~\cite{IC1981}, where
$\bar{\gamma}$ represents the global minimum of the mean-field potential. In contrast to the well-established understanding of intrinsic triaxial deformation,
dynamical triaxialilty within the IBM appears much more involved~\cite{Castanos1984,Elliott1986,Vogel1996}, which constitutes one of the primary focuses of the present study.

For our purpose, the model Hamiltonian is constructed as follows~\cite{Teng2025}:
\begin{eqnarray}\label{H}
\hat{H}=a_1\hat{n}_d+a_2\hat{Q}^\chi\cdot\hat{Q}^\chi+a_3\hat{V}_3+b_1(\hat{L}\times \hat{Q}^\chi\times \hat{L})^{(0)}+b_2\hat{L}^2\,
\end{eqnarray}
where $a_{i=1,2,3}$ and $b_{k=1,2}$ denote adjustable parameters, and
\begin{eqnarray}
&&\hat{n}_d=d^\dag\cdot\tilde{d},\\
&&\hat{L}_u=\sqrt{10}(d^\dag\times\tilde{d})_u^{(1)},\\
&&\hat{Q}_u^\chi=(d^\dag\times\tilde{s}+s^\dag\times\tilde{d})_u^{(2)}+\chi(d^\dag\times\tilde{d})_u^{(2)},\\
&&\hat{V}_3=(d^\dag\times d^\dag\times d^\dag)^{(3)}\cdot(\tilde{d}\times\tilde{d}\times\tilde{d})^{(3)}\, .
\end{eqnarray}
For simplicity, the four-body term $LQQL$ is not included in the present analysis, although
this term is equally important as the three-body term $LQL$ in the rotor mapping proceeder for the IBM description of triaxial rotor modes~\cite{Smirnov2000,Zhang2014,Teng2024}.
The first two terms in Eq.~(\ref{H}) clearly correspond to the consistent-$Q$ Hamiltonian \cite{Warner1983}, and are frequently employed to describe the dynamical evolutions between different symmetry
limits (collective modes) in the IBM~\cite{Iachellobook}. These symmetry limits include U(5) (spherical vibrator), SU(3) (axially deformed rotor), and O(6) ($\gamma$-unstable rotor).
In addition, the cubic term between the $d$ bosons, $\hat{V}_3$, is incorporated into the Hamiltonian to model specific triaxiality effects~\cite{Heyde1984,Sorgunlu2008} on $B(E2)$ structures~\cite{Sorgunlu2008,Casten1985,Ramos2000I,Ramos2000II}, since its triaxiality picture has been well established at the mean-field level~\cite{IC1981,Heyde1984}. As noted in ~\cite{Heyde1984}, such a cubic term can be interpreted as a renormalized effective interaction arising from the contribution of the $g$ boson. The Hamiltonian defined in (\ref{H}), with $N$-dependent setting
for $a_{i=1,2,3}$ and $b_{k=1,2}$, has recently been applied to describe the anomalous $B(E2)$ structures in neutron-deficient Os isotopes~\cite{Teng2025}. The first three terms of (\ref{H}) was previously applied to construct the configuration-mixing IBM Hamiltonian for modeling structural evolution in Ge, Se and Kr isotopes~\cite{Nomura2017I,Nomura2017II}. It is therefore believed that the Hamiltonian form given in (\ref{H}) is sufficient to address all typical quadrupole modes in the IBM on an equal footing.

To describe $B(E2)$ transitions and quadrupole moments, the $E2$ operator is chosen as
\begin{eqnarray}\label{E2}
T(E2)=e\hat{Q}^\chi\, ,\end{eqnarray}
where $e$ represents the effective charge, and the $\hat{Q}^\chi$ operator is identical to the one included in the Hamiltonian (\ref{H}).
The $B(E2)$ transitions can be then evaluated via
\begin{eqnarray}
B(E2;L_i\rightarrow L_f)=\frac{|\langle\alpha_fL_f\parallel T(E2)\parallel\alpha_iL_i\rangle|^2}{2L_i+1}\,
\end{eqnarray}
with $\alpha$ representing generally all the quantum numbers except for $L$.
Similarly, the quadrupole moments can be evaluated via
\begin{equation}
Q(L)=\sqrt{\frac{16\pi}{5}}\langle\alpha LM|T(E2)|\alpha LM\rangle|_{M=L}\, .
\end{equation}

To identify the mean-field geometry, one needs to determine the classical limit of the Hamiltonian. This can be achieved
by adopting the coherent state defined as~\cite{Iachellobook}
\begin{eqnarray}\label{coherent}
|\beta, \gamma, N\rangle=G[s^\dag + \beta \mathrm{cos} \gamma~
d_0^\dag\ + \frac{1}{\sqrt{2}} \beta \mathrm{sin} \gamma (d_2^\dag +
d_{ - 2}^\dag)]^N |0\rangle\,
\end{eqnarray}
with the normalization factor given by $G=1/\sqrt{N!(1+\beta^2)^N}$.
Subsequently, the full-order classical potential function corresponding to the Hamiltonian (\ref{H}) can be derived as
\begin{eqnarray}\label{V}
V(\beta,\gamma,N)&=&\langle\beta, \gamma,
N|\hat{H}|\beta, \gamma,N\rangle\\\nonumber
&=&a_1\frac{N\beta^2}{(1+\beta^2)}+a_2\Big[N\frac{(\beta^2+\chi^2\beta^2+5)}{1+\beta^2}\\ \nonumber
&~&+N(N-1)\frac{(4\beta^2-4\sqrt{\frac{2}{7}}\chi\beta^3\mathrm{cos}3\gamma+\frac{2}{7}\chi^2\beta^4)}{(1+\beta^2)^2}\Big]\\
\nonumber
&~&+a_3N(N-1)(N-2)\frac{(\beta^6\mathrm{cos}^23\gamma-\beta^6)}{7(1+\beta^2)^3}\\ \nonumber
&~&+b_1\Big[N(N-1)\frac{(2\sqrt{105}\chi\beta^4-14\sqrt{30}\beta^3\mathrm{cos}3\gamma)}{35(1+\beta^2)^2}\\ \nonumber
&~&+N\frac{\sqrt{105}\chi\beta^2}{5(1+\beta^2)}\Big]+b_2N\frac{6\beta^2}{1+\beta^2}\, .
\end{eqnarray}

\begin{figure*}
\begin{center}
\includegraphics[scale=0.185]{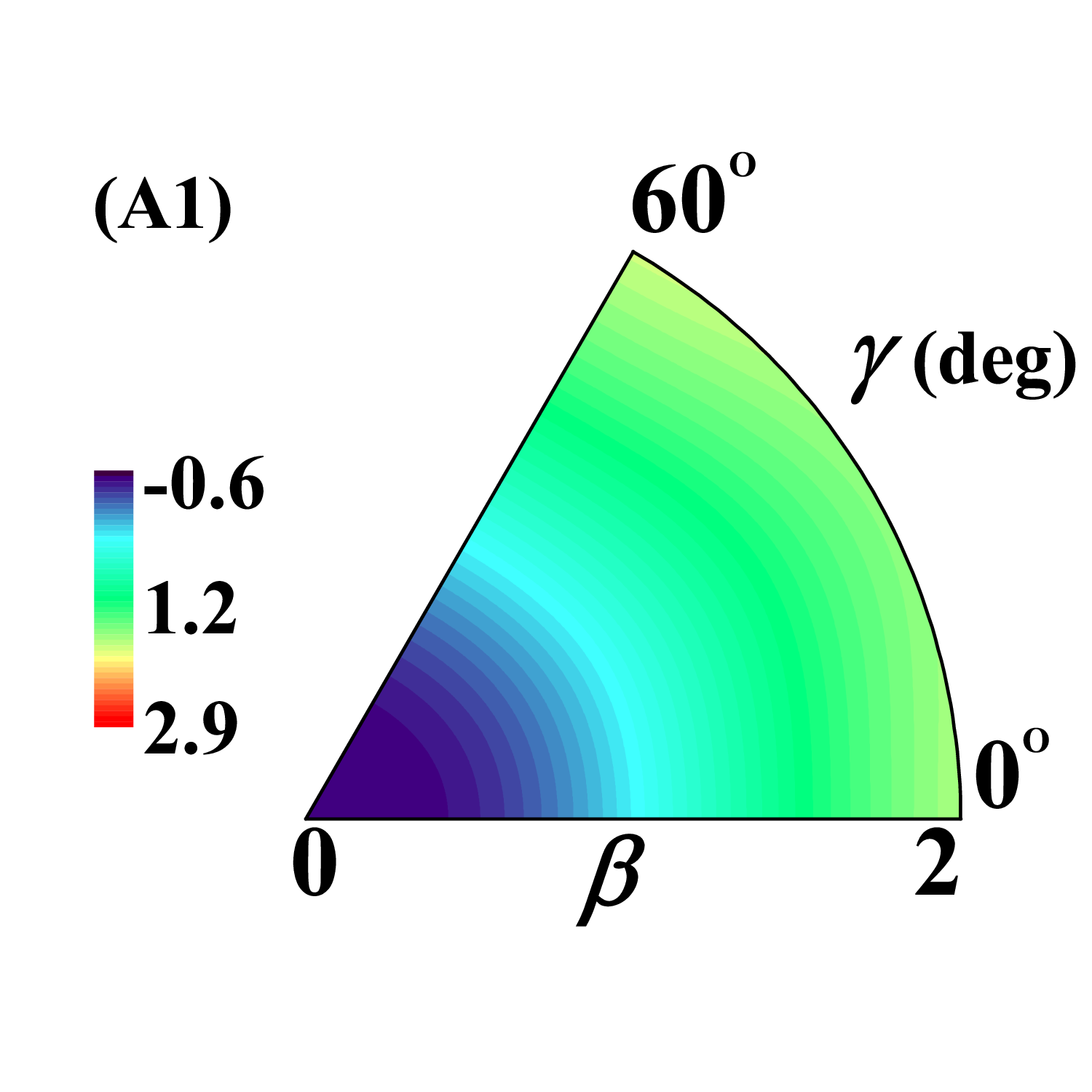}
\includegraphics[scale=0.185]{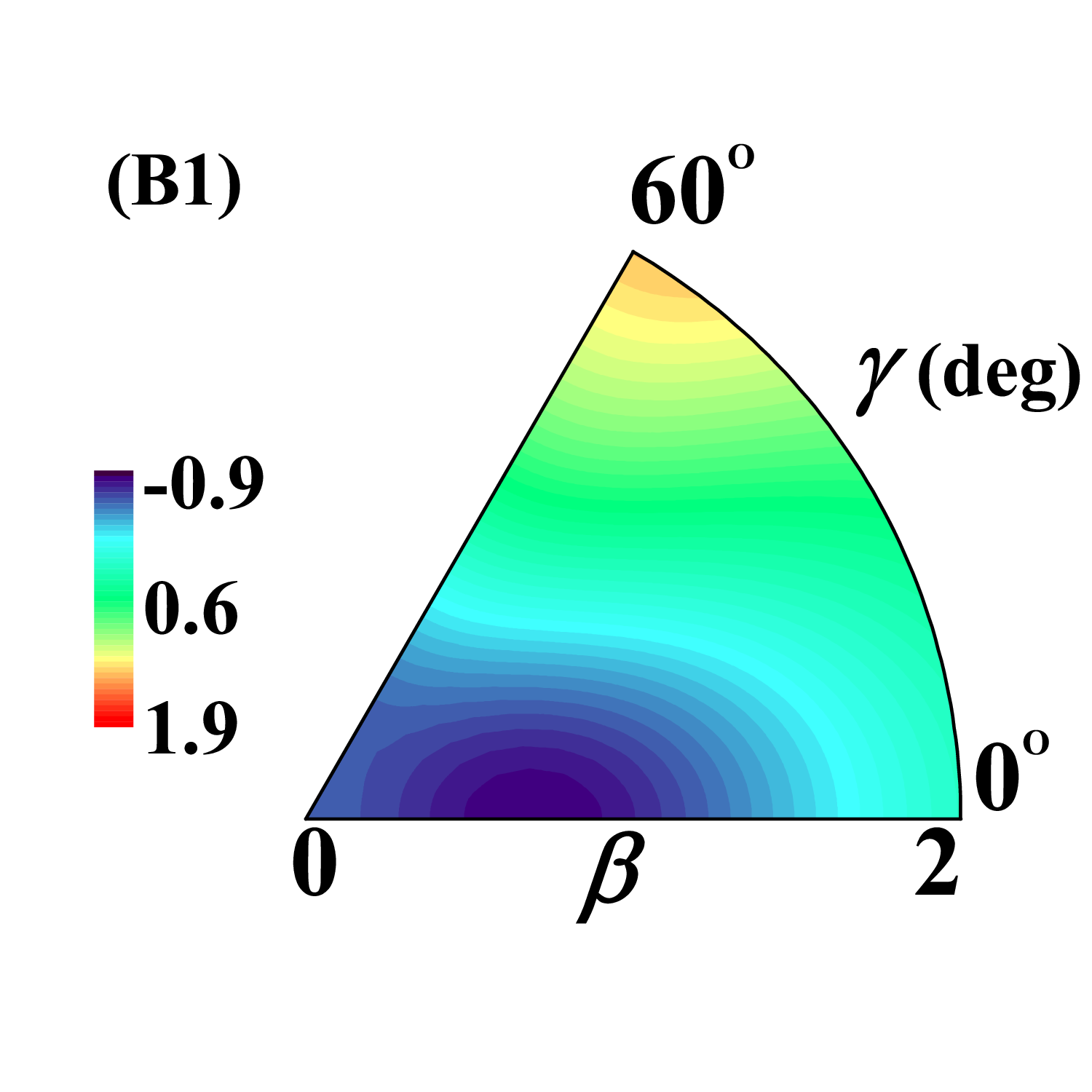}
\includegraphics[scale=0.185]{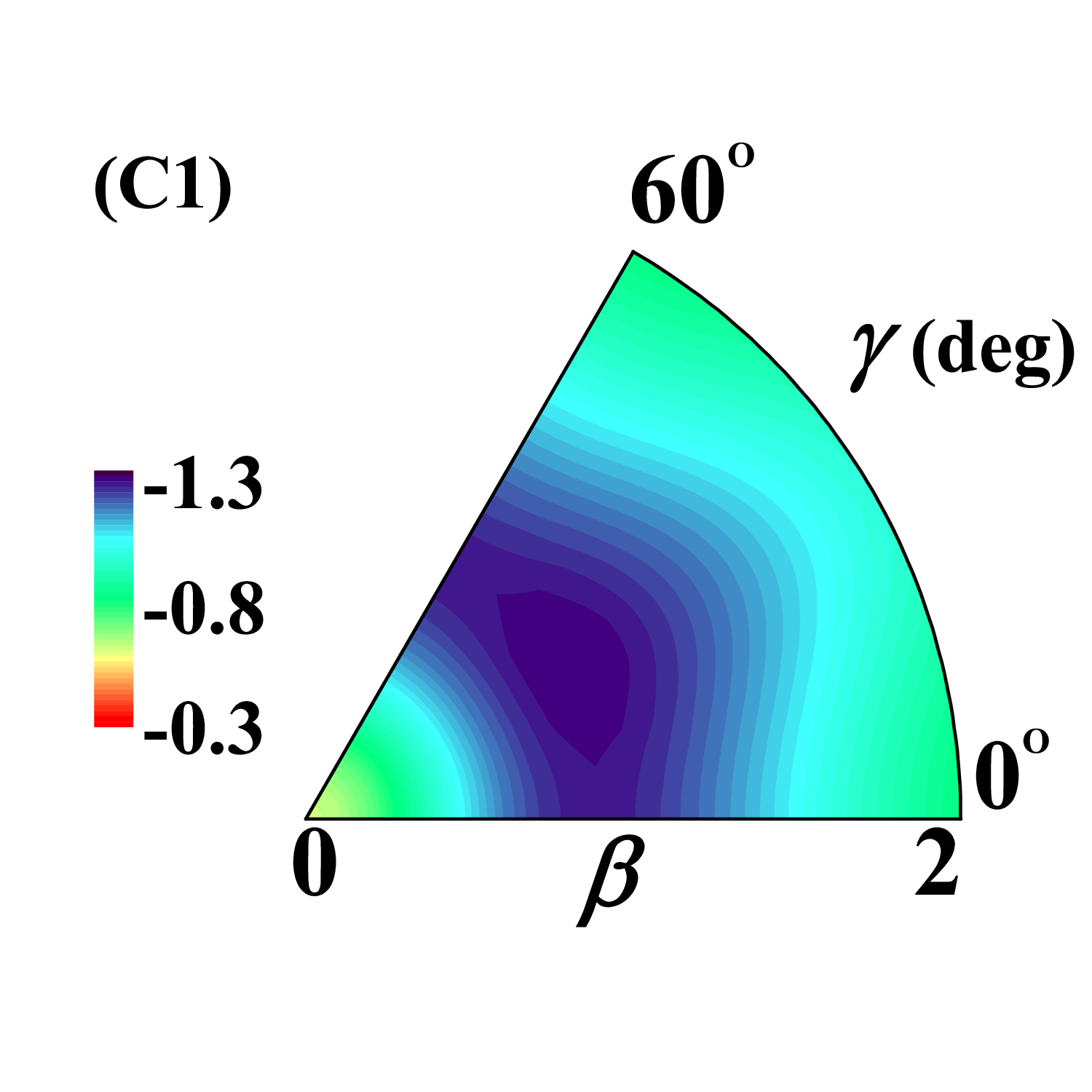}
\includegraphics[scale=0.18]{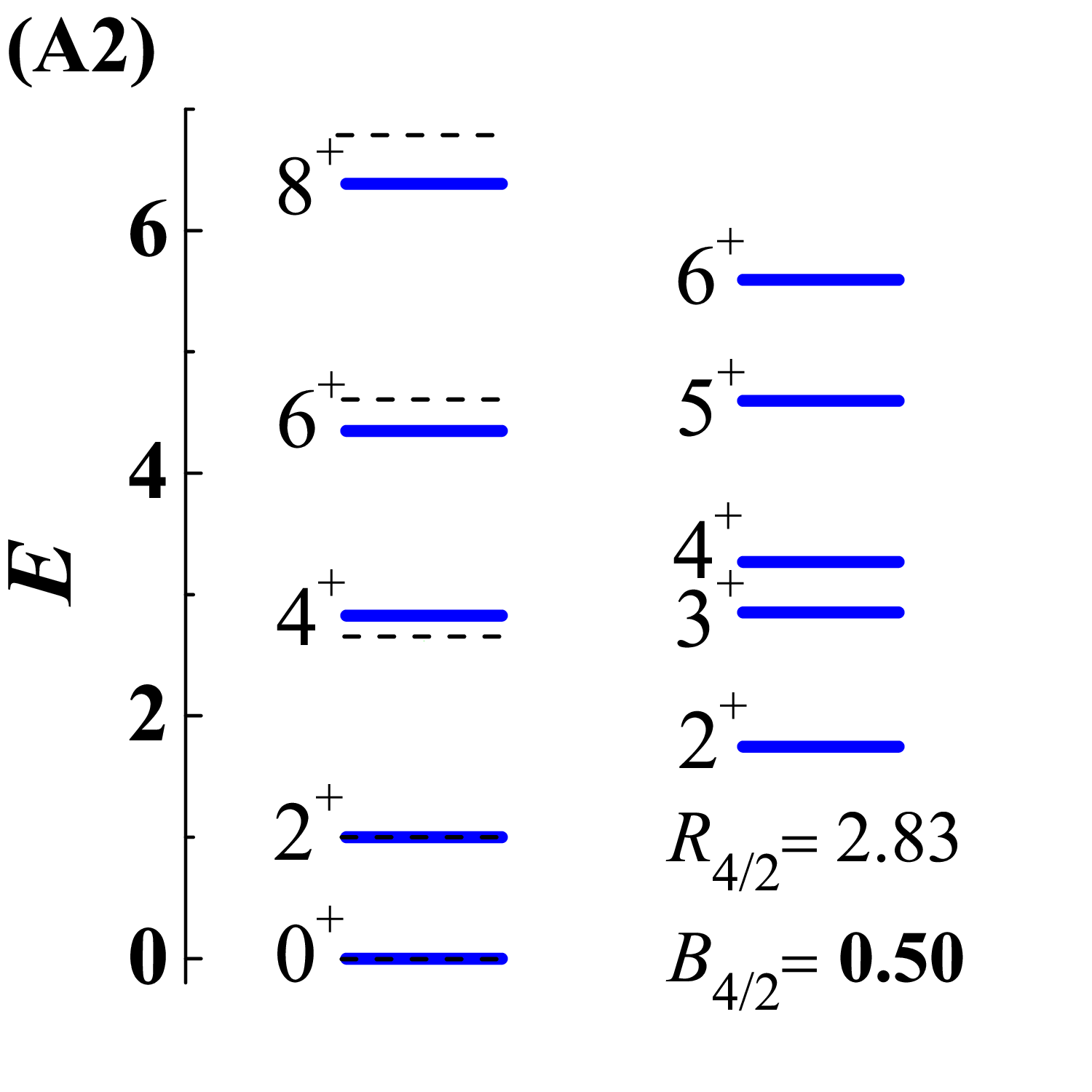}
\includegraphics[scale=0.18]{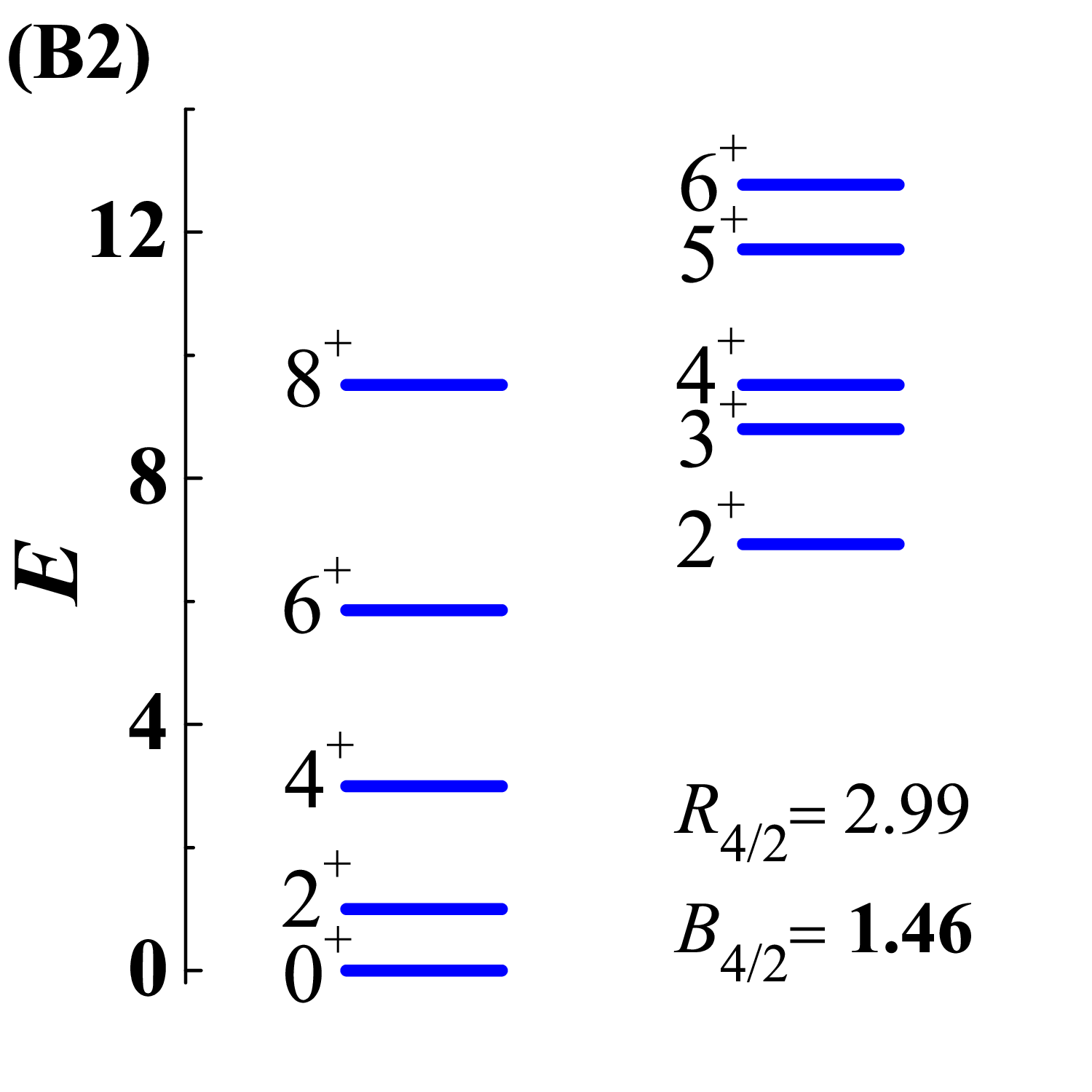}
\includegraphics[scale=0.18]{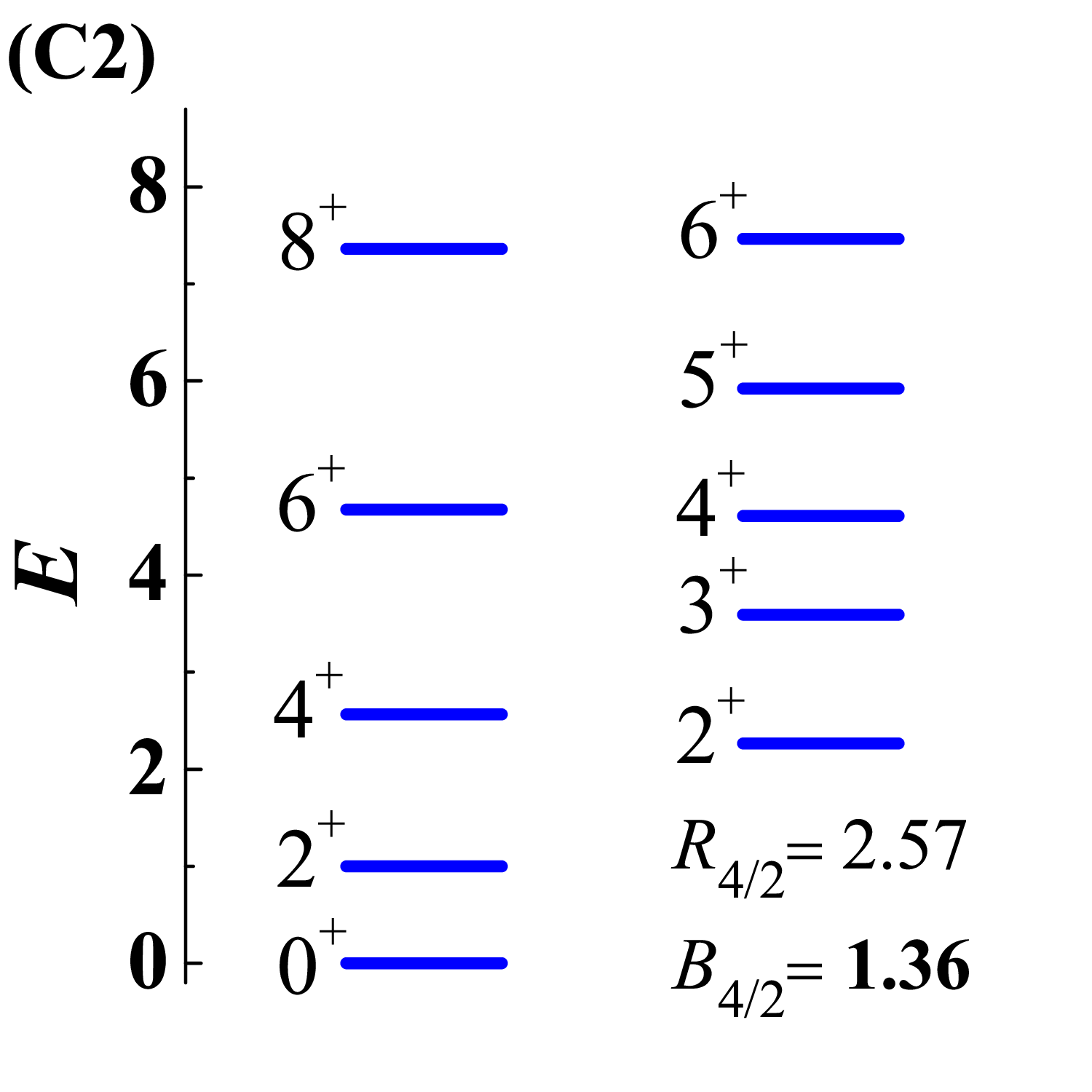}
\caption{(A1) Contour plot for the potential function ($N=9$) described by (\ref{V}) with all parameters (detailed in the text) adopted from \cite{Teng2025}. (B1) The same as in (A1), but obtained by setting $b_1=b_2=0$. (C1) The same in (A1), but obtained with $a_1=b_1=b_2=\chi=0$. (A2) The level pattern (normalized to $E(2_1^+)=1.0$) corresponding to the case in (A1), where the dashed short lines denote the values extracted from the data for $^{170}$Os. (B2) The level pattern corresponding to the case in (B1). (C2) The level pattern corresponding to the case in (C1).}\label{F1}
\end{center}
\end{figure*}

The prior analyses~\cite{Zhang2024,Teng2025} indicate that the $B(E2)$ anomaly may prefer cases involving a mixture of different collective modes including triaxial rotor.
Certainly, an ideal triaxial (rigid) rotor suggests intrinsic $\gamma$ deformation at the mean-field level. Nonetheless, when other collective modes are involved, as is expected in realistic situation,
the resulting geometry may evolve into a $\gamma$-softer state. To illustrate this point, we present an example with the parameters being chosen to match those obtained by fitting the data for $^{170}$Os as reported in \cite{Teng2025}. In that study, the same Hamiltonian was used, but with the parameters expressed in an $N$-dependent form, which demonstrates that rotor-like terms, $\langle LQL\rangle$ and $\langle L^2\rangle$, contribute negligibly to the ground-state potential surface in the large-$N$ limit.
Specifically, the corresponding parameters are adopted (in MeV) as $a_1=0.25,~a_2=-0.115/N,~a_3=0.368/N^2,~b_1=-0.902/N^2$, and $b_2=0.127/N$ along with the boson number $N=9$ and the dimensionless parameter $\chi=-0.98$. As demonstrated in \cite{Teng2025}, these parameters allow for a good reproduction of the yrast level energies and $B(E2)$ anomaly feature in $^{170}$Os~\cite{Goasduff2019}. In this work, we do not aim to achieve a better fit to the limited available data but instead utilize these parameters to model a $B(E2)$ anomaly system for theoretical analysis.

To examine the influence of different terms on the mean-field picture, contour plots of the potential function corresponding to the Hamiltonian are presented in Fig.~\ref{F1}, with the parameter assignment through three steps. This is achieved by retaining only the relevant terms and setting the irrelevant parameters to zero at each step. As show in Fig.~\ref{F1}(A1), the potential surface $V(\beta,\gamma,N)$ exhibits a spherical-like deformation. Although $\gamma$ softness or even $\gamma$ instability can be clearly observed, it is difficult to associate this mean-field picture with the dynamical structure characterized by $R_{4/2}=2.83$, $B_{4/2}=0.50$ and $Q(2_1^+)/e\simeq-8.5$, which is generated by the Hamiltonian under the given parameters, as illustrated in Fig.~\ref{F1}(A2). In addition, a low-energy yrare band is predicted to emerge in this case, which is typically a signature of triaxiality and potentially related to band mixing~\cite{Wood2004,Allmond2008}. This point can alternatively be understood from the calculated $B(E2)$ ratios, including $\frac{B(E2;2_2^+\rightarrow0_1^+)}{B(E2;2_1^+\rightarrow0_1^+)}=0.002$, $\frac{B(E2;2_2^+\rightarrow2_1^+)}{B(E2;2_1^+\rightarrow0_1^+)}=0.49$, $\frac{B(E2;4_1^+\rightarrow2_2^+)}{B(E2;2_1^+\rightarrow0_1^+)}=0.18$, $\frac{B(E2;4_2^+\rightarrow2_1^+)}{B(E2;2_1^+\rightarrow0_1^+)}=0.77$ and $\frac{B(E2;4_2^+\rightarrow2_2^+)}{B(E2;2_1^+\rightarrow0_1^+)}=0.07$.
These results clearly demonstrate that there is a strong mixing between the yrast band and yrare band. Note that the other excited bands may
also contribute to the band-mixing picture, but this does not alter the conclusion.
As further seen in Fig.~\ref{F1}(B1), where the contributions from the rotor-like terms (collective terms) are neglected, the combination of the first three terms (intrinsic terms) in the Hamiltonian (\ref{H}) may instead give rise to an axially deformed potential with a certain degree of softness in both $\beta$ and $\gamma$. Accordingly, the Hamiltonian generates a level pattern with $R_{4/2}=2.99$, $B_{4/2}=1.46$ and $Q(2_1^+)/e\simeq-10.2$, as shown Fig.~\ref{F1}(B2). The other relevant $B(E2)$ ratios in this case are $\frac{B(E2;2_2^+\rightarrow0_1^+)}{B(E2;2_1^+\rightarrow0_1^+)}=0.01$, $\frac{B(E2;2_2^+\rightarrow2_1^+)}{B(E2;2_1^+\rightarrow0_1^+)}=0.12$, $\frac{B(E2;4_1^+\rightarrow2_2^+)}{B(E2;2_1^+\rightarrow0_1^+)}=0.01$, $\frac{B(E2;4_2^+\rightarrow2_1^+)}{B(E2;2_1^+\rightarrow0_1^+)}=0.001$ and $\frac{B(E2;4_2^+\rightarrow2_2^+)}{B(E2;2_1^+\rightarrow0_1^+)}=0.65$.
Clearly, the inter-band $E2$ transitions between yrast and yrare bands become much weaker. It is therefore inferred that the rotor-like term $LQL$ in the present model tends to drive the system from the axially deformed (shown in Fig.~\ref{F1}(B1)) to a more $\gamma$-soft one (shown in Fig.~\ref{F1}(A1)), yielding pronounced band-mixing effects.

Furthermore, by retaining the cubic term and the O(6) quadrupole term ($\chi=0$) in the Hamiltonian, a triaxial potential can be developed at the mean-field level, as illustrated in Fig.~\ref{F1}(C1). This result aligns well with the previous mean-field analysis given in \cite{Sorgunlu2008}. Similar to case A, the spectral pattern in case C also features a low-energy yrare band, along with $B_{4/2}=1.36$, $R_{4/2}=2.57$, and $Q(2_1^+)/e\simeq0.0$, as depicted in Fig.~\ref{F1}(C2). Notably, the zero quadrupole moment is consistent with the $\gamma=30^\circ$ mean-field deformation picture presented in Fig.~\ref{F1}(C1). The other $B(E2)$ ratios are calculated as: $\frac{B(E2;2_2^+\rightarrow0_1^+)}{B(E2;2_1^+\rightarrow0_1^+)}=0.00$, $\frac{B(E2;2_2^+\rightarrow2_1^+)}{B(E2;2_1^+\rightarrow0_1^+)}=1.35$, $\frac{B(E2;4_1^+\rightarrow2_2^+)}{B(E2;2_1^+\rightarrow0_1^+)}=0.00$, $\frac{B(E2;4_2^+\rightarrow2_1^+)}{B(E2;2_1^+\rightarrow0_1^+)}=0.00$, and $\frac{B(E2;4_2^+\rightarrow2_2^+)}{B(E2;2_1^+\rightarrow0_1^+)}=0.67$. These results closely match those derived from the Davydov rotor model at $\gamma=30^\circ$~\cite{Davydov1958}, which predicts $R_{4/2}=2.67,~B_{4/2}=1.39$ and $Q(2_1^+)=0$, along with $\frac{B(E2;2_2^+\rightarrow0_1^+)}{B(E2;2_1^+\rightarrow0_1^+)}=0.00$, $\frac{B(E2;2_2^+\rightarrow2_1^+)}{B(E2;2_1^+\rightarrow0_1^+)}=1.43$, $\frac{B(E2;4_1^+\rightarrow2_2^+)}{B(E2;2_1^+\rightarrow0_1^+)}=0.00$, $\frac{B(E2;4_2^+\rightarrow2_1^+)}{B(E2;2_1^+\rightarrow0_1^+)}=0.00$, and $\frac{B(E2;4_2^+\rightarrow2_2^+)}{B(E2;2_1^+\rightarrow0_1^+)}=0.60$. In addition, beyond developing triaxial deformation, the inclusion of the cubic term $\hat{V}_3$ may promote the emergence of the $B(E2)$ anomaly, because this term might, to a certain extent, be integrated into the normal-order expansion of the rotor-like term $LQL$, as pointed out in \cite{Berghe1985}.

Overall, the results given in Fig.~\ref{F1} suggest that the mean-field calculations can offer only qualitative insights into the relationship between triaxiality and band mixing in a $B(E2)$ anomaly system. A more precise indicator of triaxiality should be derived from wave functions obtained by solving from the Hamiltonian, especially in cases exhibiting considerable $\gamma$ softness. To achieve this, we will further investigate the so-called effective $\gamma$ deformations in yrast states and their fluctuations.

\begin{table}
\caption{The calculated effective $\gamma$ deformations (in degrees) and their fluctuations in yrast states based on the Eq.~(\ref{effgamma})-Eq.~(\ref{delta}),
where $\gamma^{i}$ ($i$=A, B, C) represent the results obtained by using the parameters from the correspond panels in Fig.~\ref{F1}}
\begin{center}
\label{T1}
\begin{tabular}{ccc|cc|cc}\hline\hline
$L^\pi$ &$\gamma_\mathrm{a}^{\mathrm{A}}$&$\gamma_\mathrm{b}^{\mathrm{A}}~~(\Delta\gamma_\mathrm{b}^{\mathrm{A}})$&$\gamma_\mathrm{a}^{\mathrm{B}}$&$\gamma_\mathrm{b}^{\mathrm{B}}~~(\Delta\gamma_\mathrm{b}^{\mathrm{B}})$
&$\gamma_\mathrm{a}^{\mathrm{C}}$
&$\gamma_\mathrm{b}^{\mathrm{C}}~~(\Delta\gamma_\mathrm{b}^{\mathrm{C}})$\\ \hline
$0_1^+$&$10^\circ$&$9^\circ~~(8^\circ)$&$10^\circ$&$9^\circ~~(8^\circ)$&$30^\circ$&$12^\circ~~(11^\circ)$\\
$2_1^+$&$10^\circ$&$9^\circ~~(8^\circ)$&$9^\circ$&$7^\circ~~(6^\circ)$&$30^\circ$&$12^\circ~~(11^\circ)$\\
$4_1^+$&$20^\circ$&$23^\circ~(12^\circ)$&$8^\circ$&$6^\circ~~(5^\circ)$&$30^\circ$&$12^\circ~~(11^\circ)$\\
$6_1^+$&$27^\circ$&$34^\circ~~(8^\circ)$&$7^\circ$&$6^\circ~~(5^\circ)$&$30^\circ$&$12^\circ~~(10^\circ)$\\
$8_1^+$&$31^\circ$&$44^\circ~~(6^\circ)$&$7^\circ$&$5^\circ~(4^\circ)$&$30^\circ$&$11^\circ~~~~(9^\circ)$\\
\hline\hline
\end{tabular}
\end{center}
\end{table}

The effective $\gamma$ deformation can be determined using the definition provided in \cite{Elliott1986}:
\begin{eqnarray}\label{effgamma}
\mathrm{Cos}(3\gamma_\mathrm{a})=-\Big(\frac{7}{2\sqrt{5}}\Big)^{1/2}\frac{\langle(\hat{Q}^\chi\times\hat{Q}^\chi\times\hat{Q}^\chi)^{(0)}\rangle}{(\langle\hat{Q}^\chi\times\hat{Q}^\chi)^{(0)}\rangle^{3/2}}\, ,
\end{eqnarray}
where $\langle\rangle$ represents the expectation value of the corresponding operator under the given state.
Triaxiality inferred from Eq.~(\ref{effgamma}) represents a dynamic result rather than one arising from the intrinsic $\gamma$ deformation described by Eq.~(\ref{V}). Therefore, the effective triaxial deformation defined here as $0^\circ<\gamma_\mathrm{a}<60^\circ$ based on Eq.~(\ref{effgamma}) is referred to as dynamical triaxial deformation. It has been demonstrated~\cite{Vogel1996} that this definition  can be applied to signify triaxialilty in $\gamma$-soft systems, such as Xe and Ba nuclei with $A\sim130$. Since dynamical triaxial deformation, as discussed here, has the same meaning as effective triaxial deformation, these terms will be used interchangeably throughout this work. In the SU(3) limit ($\chi=-\sqrt{7}/2$), Eq.~(\ref{effgamma}) is consistent with another very useful formula related to $\gamma$ deformation, which is defined as~\cite{Leschber1987}
\begin{equation}\label{su3gamma}
\gamma_\mathrm{b}=\mathrm{tan}^{-1}\Big(\frac{\sqrt{3}(\mu+1)}{2\lambda+\mu+3}\Big)\, ,
\end{equation}
where $\lambda$ and $\mu$ are the quantum number used to label the SU(3) irreducible representations. This formula is very convenient for evaluating
both the average value of effective $\gamma$ deformation, $\gamma_\mathrm{b}$, and its fluctuation~\cite{Pan2003},
\begin{eqnarray}\label{delta}
\Delta\gamma_\mathrm{b}=\sqrt{\langle(\gamma_\mathrm{b}-\overline{\gamma}_\mathrm{b})^2\rangle}\, .
\end{eqnarray}
These evaluations can be performed after expanding the eigenstate in terms of the SU(3)$\supset$SO(3) basis vectors. In the following, we apply both Eq.~(\ref{effgamma}) and Eq.~(\ref{su3gamma}) to calculate the
effective $\gamma$ deformation, but only Eq.~(\ref{delta}) is used to estimate the $\gamma$ fluctuation, $\Delta\gamma_\mathrm{b}$. Using the parameters adopted in Fig.~\ref{F1},
the effective $\gamma$ deformations and $\gamma$ fluctuations are calculated for yrast states up to $L=8$, and the results are listed in Table~\ref{T1}.

As shown in Table~\ref{T1}, identical values are obtained for the ground states solved from cases A and B, showing nonzero effective $\gamma$ deformation and $\gamma$ fluctuation. This result can be readily understood, as the wave functions for all $0^+$ states (including the ground state) remain unaffected by the rotor-like terms involving $L$. This stands in sharp contrast to the dramatic differences between the two cases observed at the mean-field level given in Fig.~\ref{F1}. With the increase of spin, the effective $\gamma$ values and their fluctuations
in case A become significantly larger than those in case B, suggesting the stronger band mixing and the larger dynamical triaxiality in the former.
Particularly, the abrupt enhancement of the effective $\gamma$ deformation from $\gamma^\mathrm{A}\approx10^\circ$ to $\gamma^\mathrm{A}>20^\circ$ around $L=4$ is accompanied by the occurrence of $B(E2)$ anomaly in the present model when
the rotor-like terms are included in the calculations. Clearly, these features associated with dynamical triaxility cannot be identified solely based on the mean-field calculations.

It is worth noting that the nonzero effective $\gamma$ deformation for the ground state obtained in case A, with parameters determined by fitting data for $^{170}$Os, is qualitatively consistent with the result ($\gamma\approx15^\circ$) derived using the microscopic variation after particle number projection (PN-VAP) method as reported in \cite{Goasduff2019}, as well as the calculation based on the covariant density functional theory with the point-coupling interaction PC-PK1~\cite{Zhao2010}, which yields a potential surface with $\gamma\approx10^\circ$ for this nucleus. In fact, if the angular momentum projection method proposed in \cite{Dobes1985} is applied to perform mean-field calculations for the IBM, the resulting the ground-state ($L=0$) $\gamma$ deformation yields $\gamma\approx10^\circ$ for case A, which closely matches the effective $\gamma$ deformation listed in Table~\ref{T1}. Although angular momentum projection can also be applied to compute the $\gamma$ deformations for excited states, the calculations for $L\neq0$ become considerable more intricate due to $K$ mixing and the non-orthogonality of the coherent state wave function, and will therefore be discussed elsewhere. Certainly, different mean-field approaches can result in quantitatively distinct values of $\gamma$ deformation. Nevertheless, it is generally anticipated that systems with much smaller $R_{4/2}$ values, such as neutron-deficient nuclei~\cite{Grahn2016,Saygi2017,Cederwall2018,Goasduff2019,Zhang2021} displaying $B_{4/2}<1.0$ alongside $R_{4/2}<2.7$, will exhibit some degree of triaxial deformation or $\gamma$-softness, particularly when compared to axial rotor with $R_{4/2}>3.0$. Therefore, a meaningful assessment of the current theoretical perspective involves finding a method to determine the effective $\gamma$ deformation from the experimental observables~\cite{Werner2005}.

As further observed in the Table~\ref{T1}, in contrast to cases A and B, where $\gamma_\mathrm{a}$ and $\gamma_\mathrm{b}$ yield similar results, the effective $\gamma$ deformations in case C consistently exhibit, via Eq.~(\ref{effgamma}), maximal triaxiality with $\gamma_\mathrm{a}^\mathrm{C}=30^\circ$, for all yrast sates. While, Eq.~(\ref{su3gamma}) generates much smaller effective $\gamma$ deformations but with considerable large $\gamma$ fluctuations in comparison with the case B. It appears that in the present case, $\gamma_\mathrm{a}^\mathrm{C}=30^\circ$ aligns better with the mean-field result shown in Fig.~\ref{F1}(C1). This point can be understood from the fact that Eq.~(\ref{effgamma}), used to extract $\gamma_\mathrm{a}$, consistently employ the same $\hat{Q}^\chi$ operator as the one in the Hamiltonian. In contrast, Eq.~(\ref{su3gamma}), where the $(\lambda,\mu)$ values can be determined from the expectation value of the SU(3) generator $\hat{Q}^{\chi=-\sqrt{7}/2}$, implies that the $\hat{Q}$ operator involved in extracting $\gamma_\mathrm{b}$ differs significantly from that in the Hamiltonian, particularly for cases where $|\chi|\ll1.32$. Consequently, the effective $\gamma$ values obtained from the two methods are similar in cases A and B with $\chi=-0.98$, but diverge in case C with $\chi=0$.
It is worth mentioning that the recent analysis in \cite{Zhang2025}, which employs the extended consistent-$Q$ Hamiltonian~\cite{Fortunato2011} (similar to that defined in (\ref{H}) but with the cubic term $\hat{V}_3$ replaced by $(\hat{Q}^\chi\times\hat{Q}^\chi\times\hat{Q}^\chi)^{(0)}$), assumes a larger effective $\gamma$ deformation with $\gamma_a\approx40^\circ$ for the ground state of $^{170}$Os~\cite{Goasduff2019} when describing the experimental data. However, there exists a parameter relation, $\chi\rightarrow-\chi$ and $\hat{Q}^\chi\rightarrow-\hat{Q}^\chi$, for the adopted Hamiltonian, which preserves all calculated results unchanged except for alternatively generating the effective $\gamma$ deformation and quadrupole moment as $\gamma_\mathrm{a}\rightarrow60^\circ-\gamma_\mathrm{a}$ and $Q(L)\rightarrow-Q(L)$. Such a parameter relation also holds for the Hamiltonian given in (\ref{H}), implying that additional constraint for the parameters should account for the quadrupole moments. Considering this fact and assuming $Q(2_1^+)<0$, calculations using the parameters adopted in \cite{Zhang2025} may yield $\gamma_\mathrm{a}\approx18.7^\circ,~19.7^\circ$, and $27.7^\circ$ for $0_1^+,~2_1^+$, and $4_1^+$, respectively, along with $Q(2_1^+)/e\approx-3.1$. These results qualitatively agree with the above calculations using the Hamiltonian in (\ref{H}). Particularly, the calculations indicate that there is an approximate $10^\circ$ variation in the $\gamma_a$ values occurring around $L=4$ in this case. It is thus inferred that a dramatic change in the effective $\gamma$ deformation with increase nuclear spin may be a universal feature in the $B(E2)$ anomaly systems modeled by an IBM Hamiltonian involving the rotor-like term $LQL$ or $LQQL$~\cite{Zhang2025}.

\begin{center}
\vskip.2cm\textbf{III. Configuration Mixing}
\end{center}\vskip.2cm

In addition to the light Os nuclei, the $B(E2)$ anomaly phenomenon has also been identified in Pt isotopes with $A\approx170$.
Specifically, $^{172}$Pt was recently observed~\cite{Cederwall2018} to exhibit a significant depressed $B(E2)$ ratio, $B_{4/2}\simeq0.55(19)$, along with $R_{4/2}\approx2.34$. On the other hand, prior studies~\cite{Dracoulis1994,King1998,Harder1997,Morales2008,Ramos2009,Ramos2011,Ramos2014} suggest that 2p-2h excitations across the Z=82 proton closed shell may play a crucial role in shaping structures of Pt isotopes. The IBM configuration-mixing (IBM-CM) scheme~\cite{Duval1982} is frequently employed to model these 2p-2h crossing shell excitations~\cite{Heyde2011}, with the model space spanned using the U(6) representations $[N]\oplus[N+2]$.
Although there is currently no direct evidence linking $B(E2)$ anomaly to crossing shell excitation, we can investigate whether configuration mixing theoretically contributes to the occurrence of $B(E2)$ anomaly. For this purpose, the model Hamiltonian is constructed as follows~\cite{Ramos2009}:
\begin{eqnarray}\label{HCM} \nonumber
\hat{H}_{\mathrm{CM}}&=&\hat{P}_N^\dag\hat{H}_{A}\hat{P}_N+\hat{P}_{N+2}^\dag
(\hat{H}_{B}+\Delta)\hat{P}_{N+2}+\hat{P}_{N+2}^\dag\hat{H}_{\mathrm{mix}}\hat{P}_N\\
&~&+\hat{P}_{N}^\dag\hat{H}_{\mathrm{mix}}^\dag\hat{P}_{N+2}\, ,
\end{eqnarray}
where $\hat{P}_{N}$ and $\hat{P}_{N+2}$ are projection operators onto the $[N]$ boson space (0p-0h configuration) and $[N+2]$ boson space (2p-2h configuration), respectively.
In Eq.~(\ref{HCM}), $\hat{H}_i$ with $i=A,B$, representing the Hamiltonian for the configuration $A$ and $B$, respectively, take the same form as defined in (\ref{H}), and the term
$\hat{H}_{\mathrm{mix}}$ ($\hat{H}_{\mathrm{mix}}^\dag$) describes the mixing between the two configurations:
\begin{eqnarray}
\hat{H}_{\mathrm{mix}}=\omega(s^\dag\times s^\dag+d^\dag\times d^\dag)^{(0)}\, .
\end{eqnarray}
The parameter $\Delta$, representing the off-set energy between the two configurations, is introduced to describe the energy required to excite two particles from the lower shell to the upper shell.
The entire Hamiltonian can alternatively be expressed in matrix form~\cite{Gavrielov2019}:
\begin{eqnarray}\label{HCMMatrix}
\hat{H}_{\mathrm{CM}}=\left[ \begin{array}{cc} \hat{H}_{A} & \hat{H}_{\mathrm{mix}}^\dag \,   \\
 \hat{H}_{\mathrm{mix}} & \hat{H}_{B}+\Delta\,  \end{array} \right].
\end{eqnarray}
Accordingly, the eigenstates solved from $\hat{H}_{\mathrm{CM}}$ can be expressed as linear combinations of states in the two spaces $[N]$ and $[N+2]$:
\begin{eqnarray}
|\alpha, L\rangle=\xi_A|\alpha^A, L\rangle_N+\xi_B|\alpha^B, L\rangle_{N+2}\, ,
\end{eqnarray}
where $\alpha^i$ with $i=A,~B$ represent generally the other quantum numbers except for $L$, and the expansion coefficients satisfy $\xi_A^2+\xi_B^2=1$~\cite{Gavrielov2019}.
In the IBM-CM framework, the $E2$ operator can be chosen as~\cite{Ramos2009}
\begin{eqnarray}\label{E2CM}
T(E2)_{\mathrm{CM}}=e_A\hat{P}_N^\dag\hat{Q}_A^{\chi^A}\hat{P}_N+e_B\hat{P}_{N+2}^\dag\hat{Q}_B^{\chi^B}\hat{P}_{N+2}\, ,
\end{eqnarray}
where $e_i$ with $i=A,~B$ denote the effective boson charges and $\hat{Q}_i^{\chi^i}$ represent the corresponding quadrupole operators consistent with those used in the Hamiltonian.

\begin{figure*}
\begin{center}
\includegraphics[scale=0.15]{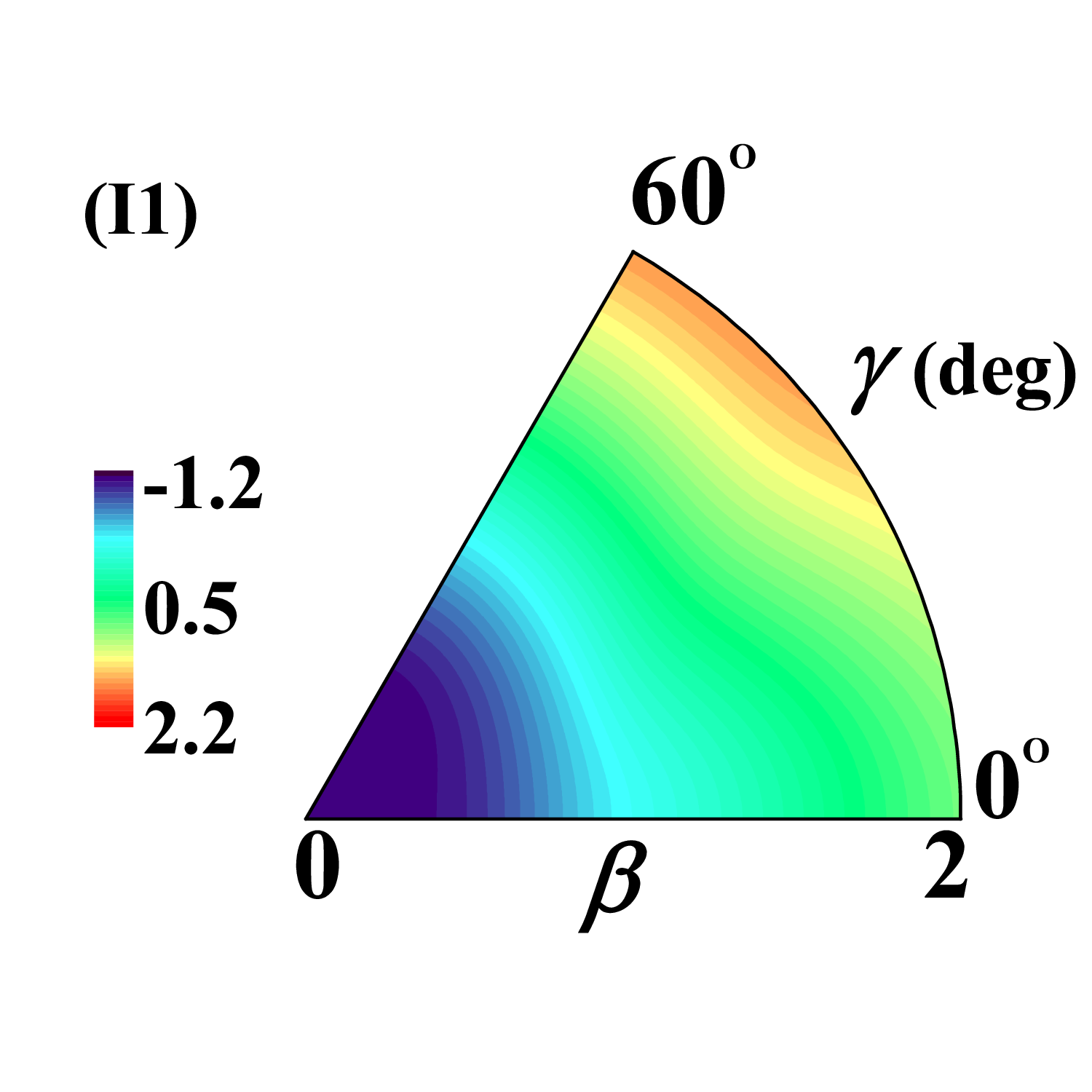}
\includegraphics[scale=0.15]{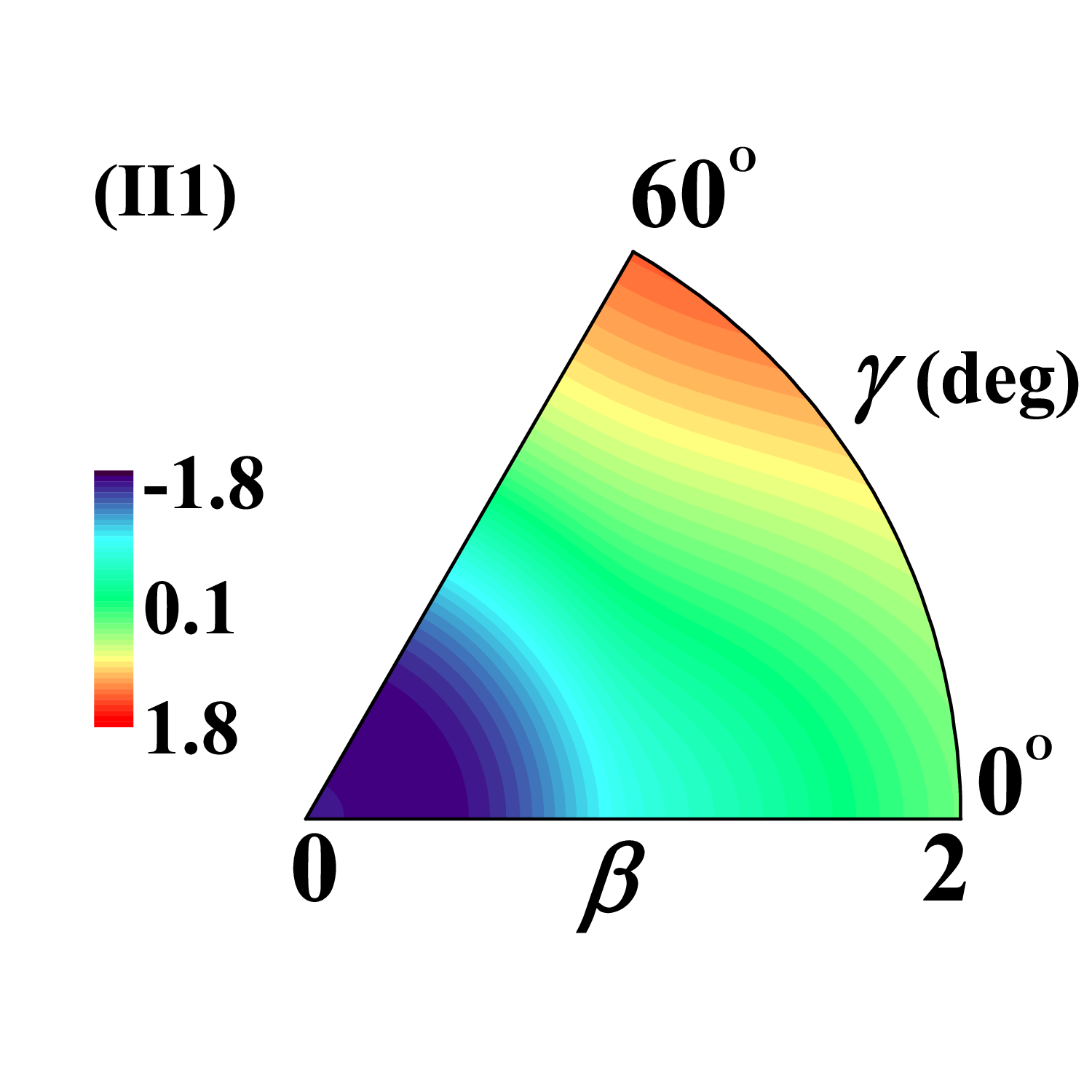}
\includegraphics[scale=0.15]{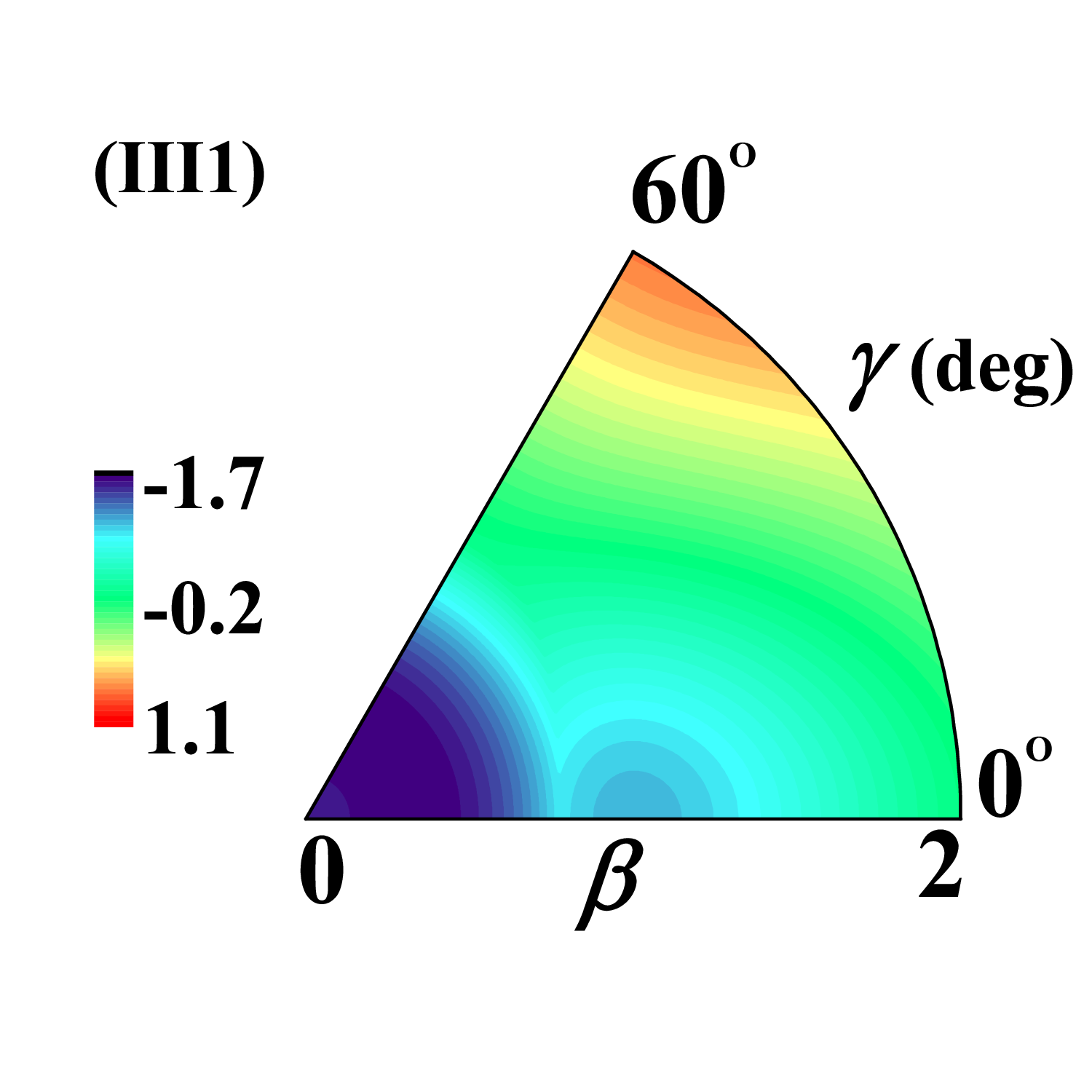}
\includegraphics[scale=0.15]{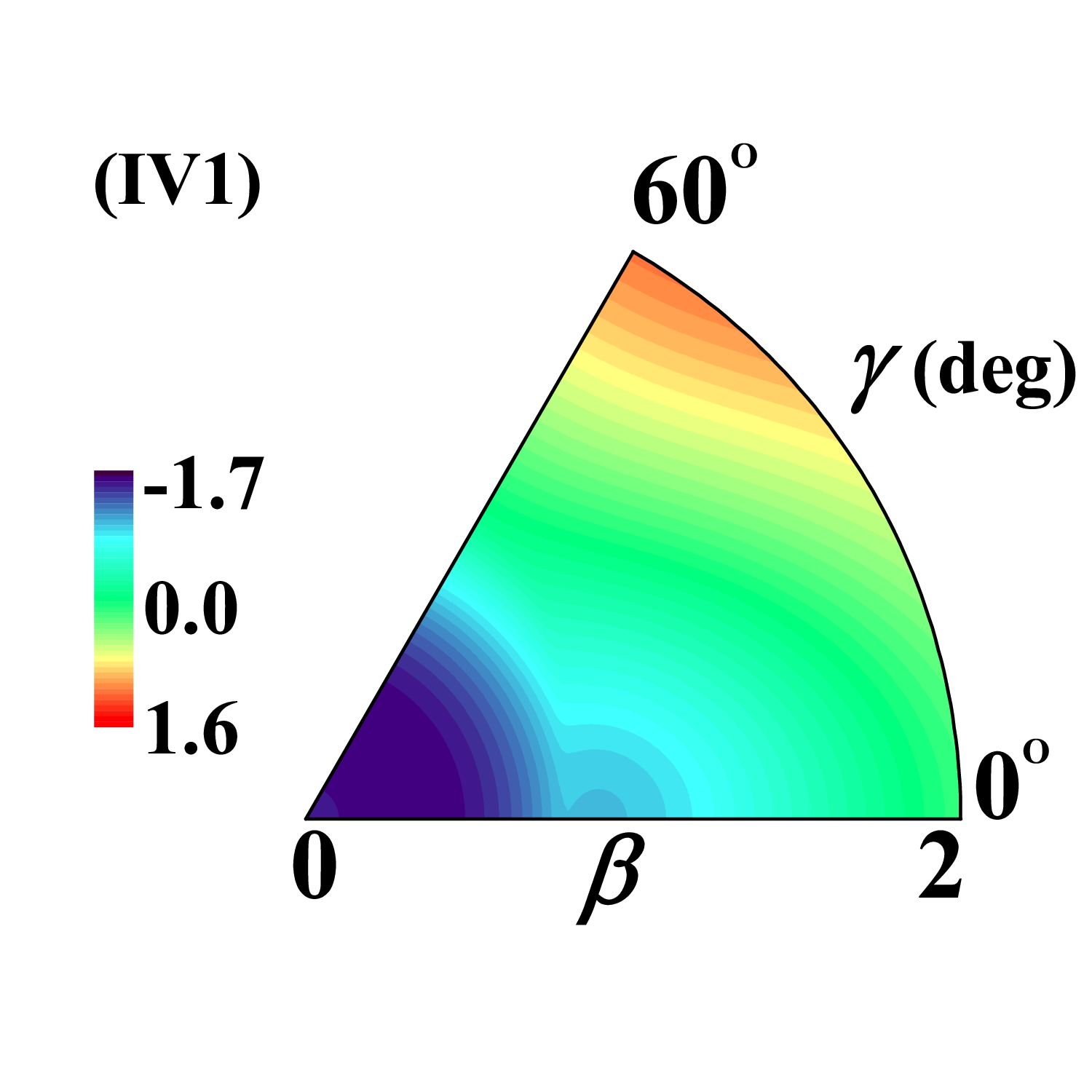}
\includegraphics[scale=0.15]{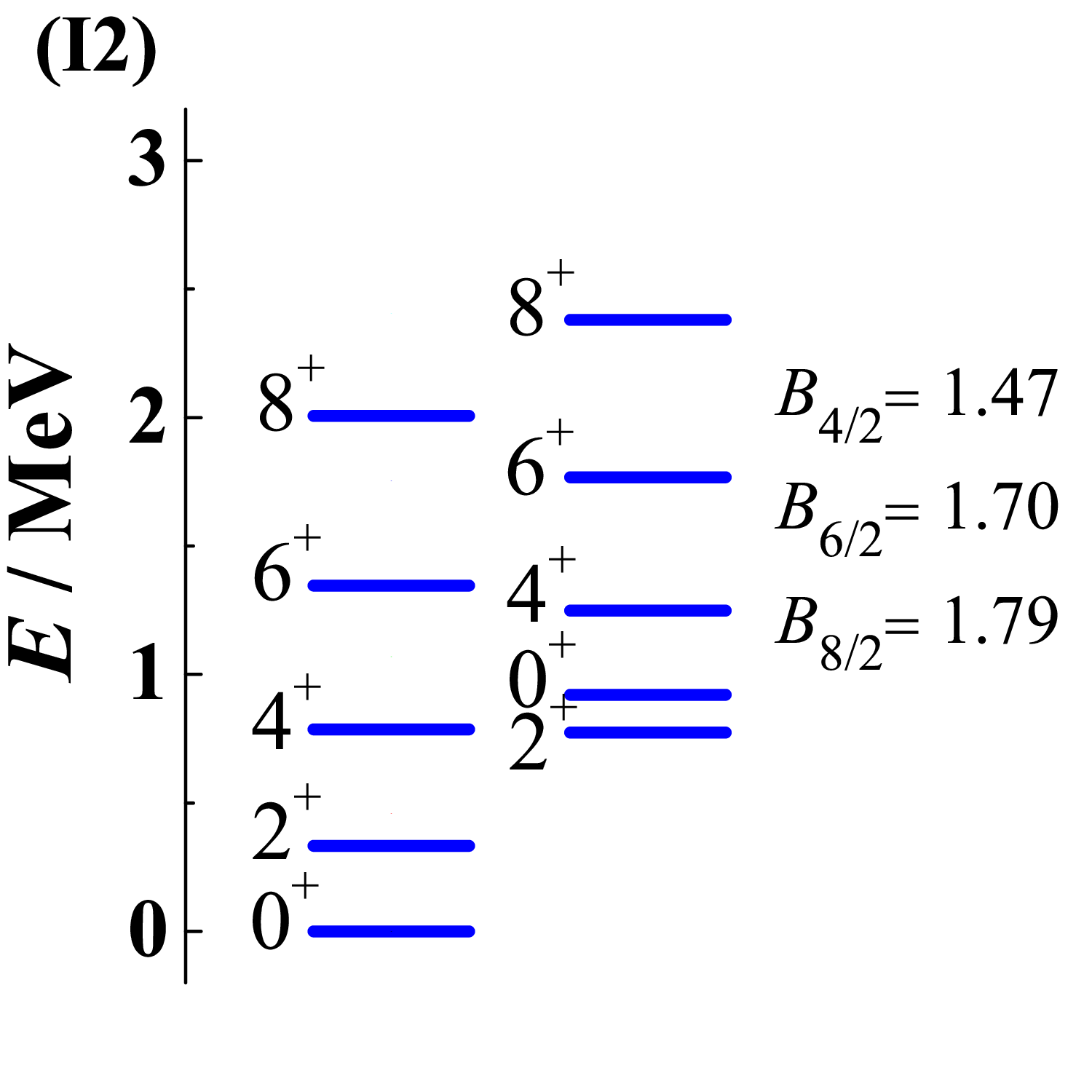}
\includegraphics[scale=0.15]{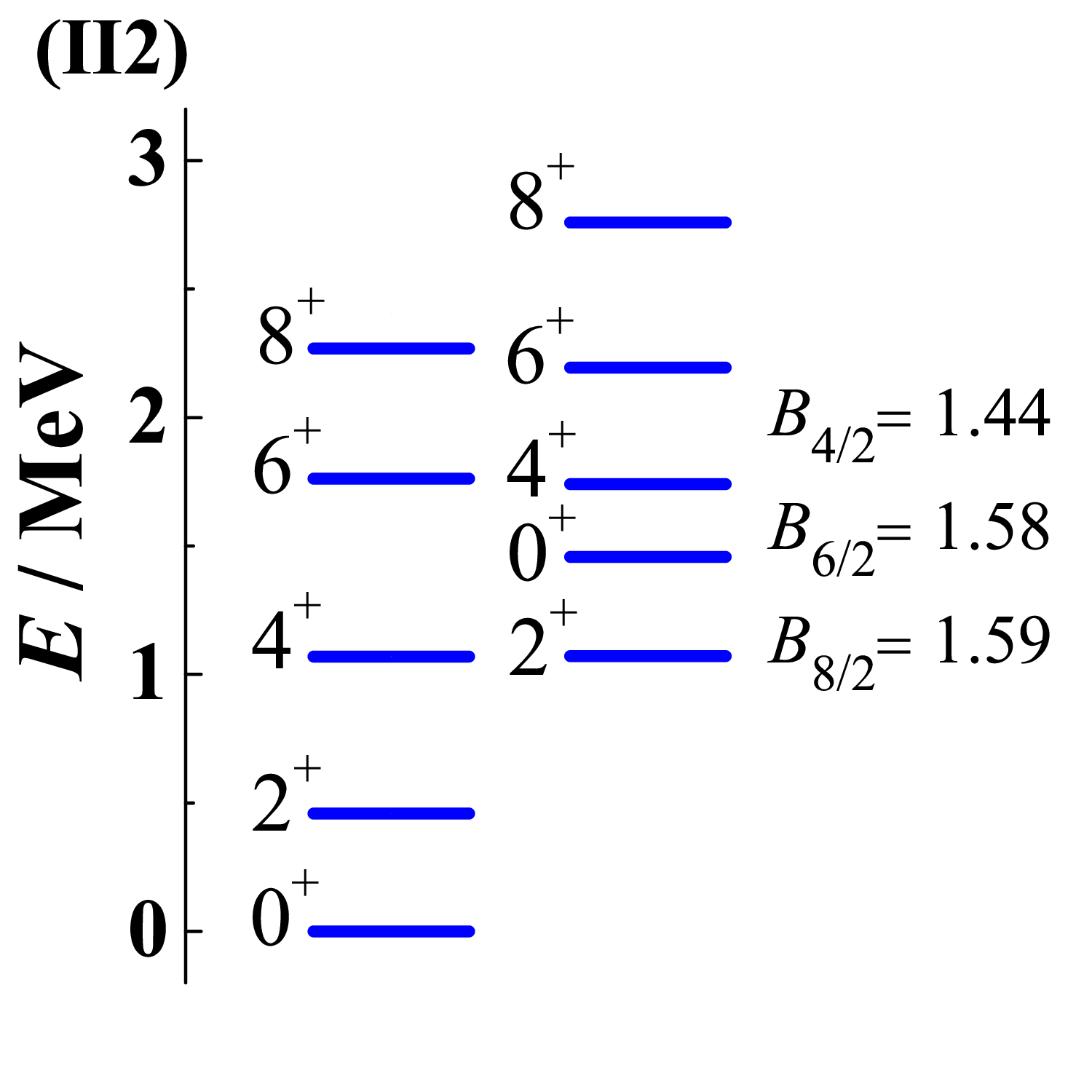}
\includegraphics[scale=0.15]{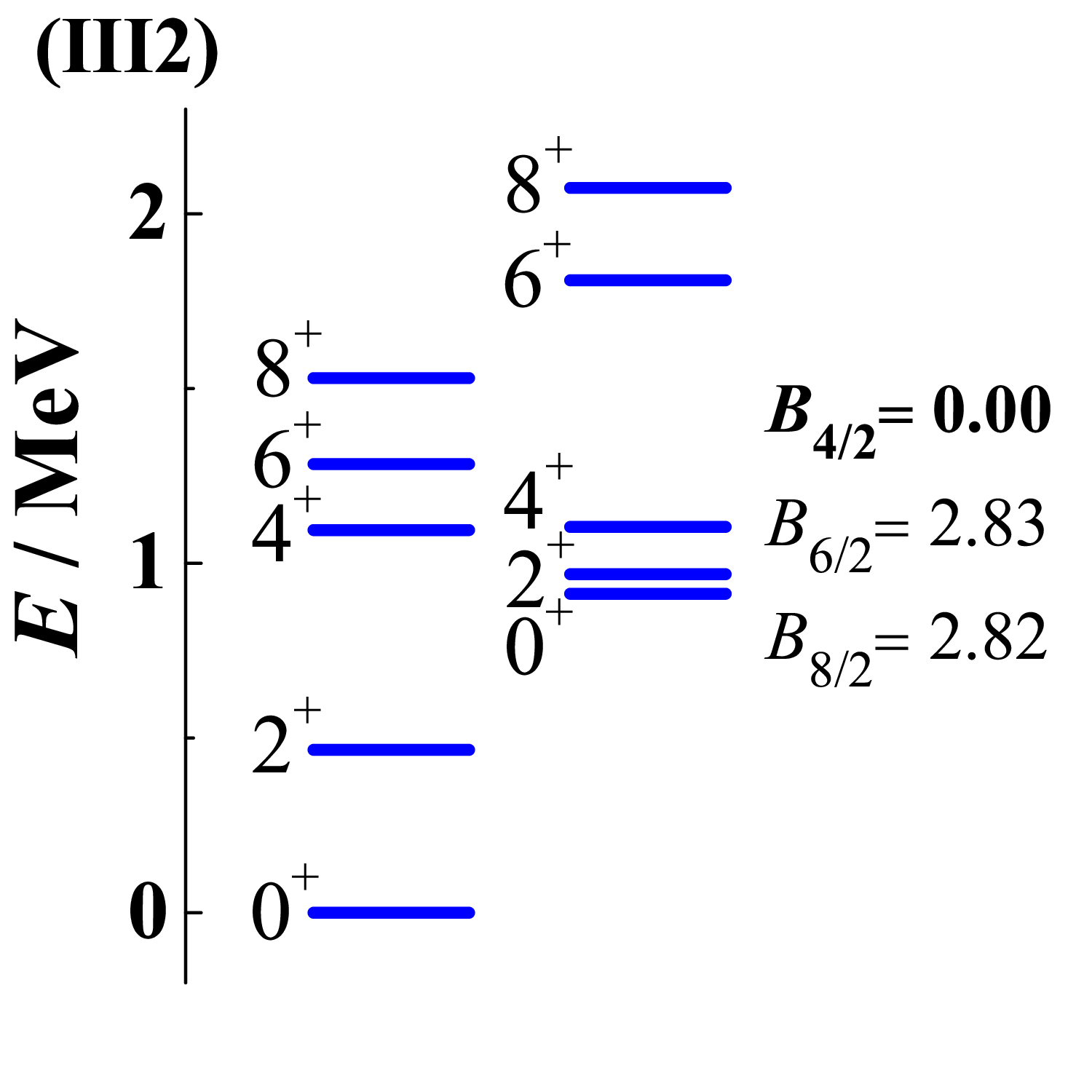}
\includegraphics[scale=0.15]{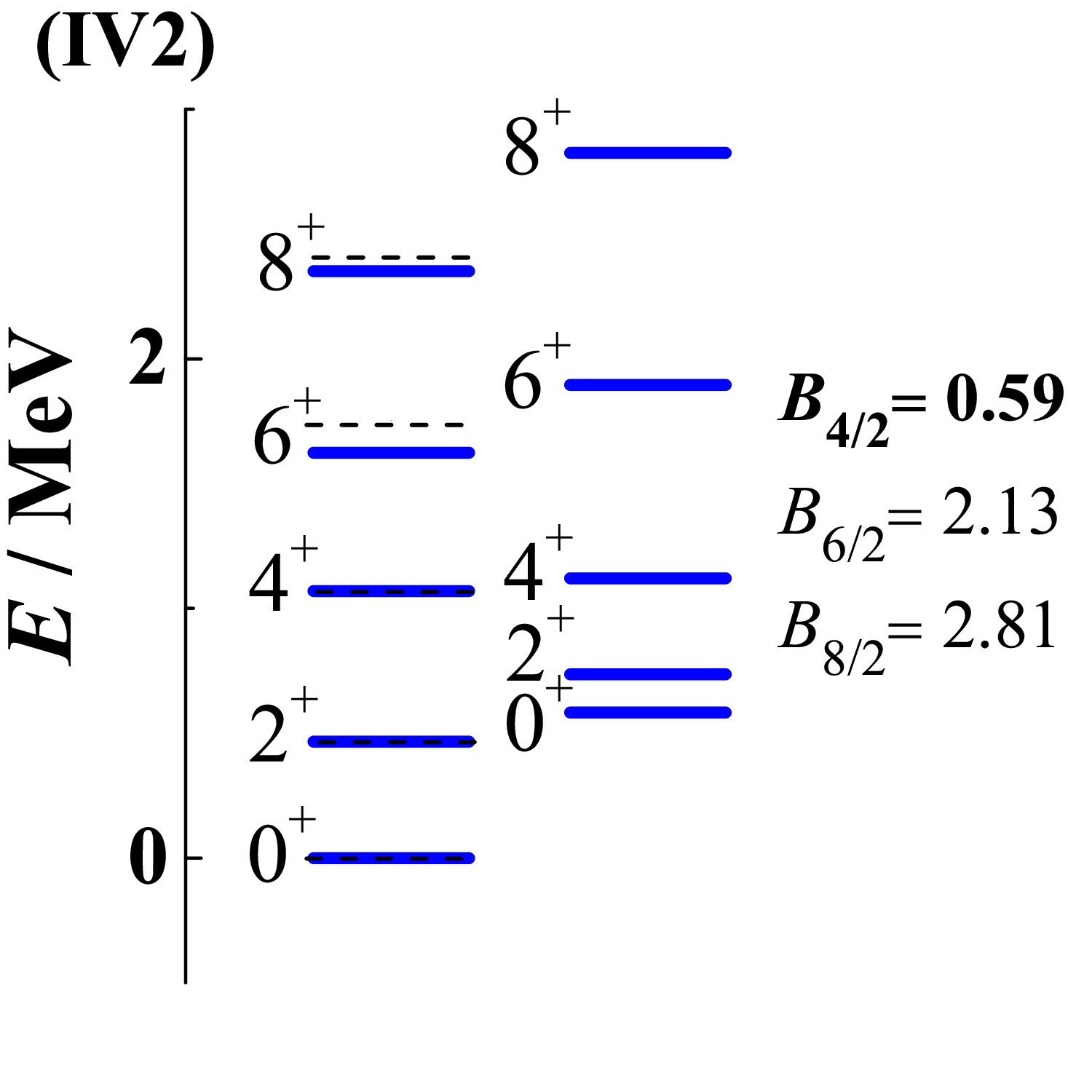}
\caption{(I1) The potential surface solved from (\ref{VMatrix}) with the parameters (in MeV) taken from \cite{Harder1997,Morales2008}. (II1) The same as in (I1) but with the parameters taken from \cite{Ramos2009}. (III1) The same in (II1) but obtained by readjusting the parameters based on those used in (II1). (IV1) The same as in (III1) but with further adjustment of the parameters. (I2) The yrast and yrare levels corresponding to the case in (I1). (II2) The level pattern corresponding to the case in (II1). (III2) The level pattern corresponding to the case in (III1). (IV2) The level pattern corresponding to the case in (IV1), where the short dashed lines denote the experimental data for $^{172}$Pt~\cite{Cederwall2018}. The parameters adopted in each cases have been detailed in the text, and the $B_{L/2}$ ratios are calculated using the $E2$ operator in (\ref{E2CM}) by assuming the effective charges $e_A=e_B$ in different cases for simplicity.}\label{F2}
\end{center}
\end{figure*}

The mean-field potential associated with an IBM-CM Hamiltonian can be derived using a matrix coherent state method~\cite{Frank2004,Frank2006}. For the Hamiltonian in (\ref{HCMMatrix}), one obtains a $2\times2$ potential energy matrix
\begin{eqnarray}\label{VMatrix}
V_{\mathrm{CM}}=\left[ \begin{array}{cc} V(\beta,\gamma,N)_A & V_{\mathrm{mix}}^{N,N+2} \,   \\
 V_{\mathrm{mix}}^{N+2,N} & V(\beta,\gamma,N+2)_B+\Delta\,  \end{array} \right]\, ,
\end{eqnarray}
where the potentials, $V(\beta,\gamma,N^\prime)_{i}\equiv\langle \beta,\gamma,N^\prime|\hat{H}_{i}|\beta,\gamma,N^\prime\rangle$ with $i=A,~B$, have the same form as that described by Eq.~(\ref{V}), and the nondiagonal matrix elements are defined as
\begin{eqnarray}
&&V_{\mathrm{mix}}^{N+2,N}=\langle \beta,\gamma,N+2\mid\hat{H}_{\mathrm{mix}}\mid\beta,\gamma,N\rangle\\
&&V_{\mathrm{mix}}^{N,N+2}=\langle \beta,\gamma,N\mid\hat{H}_{\mathrm{mix}}^\dag\mid\beta,\gamma,N+2\rangle\, .
\end{eqnarray}
By using the coherent state defined in (\ref{coherent}), one can derive
\begin{eqnarray}\label{Vmixpotential}
V_{\mathrm{mix}}^{N+2,N}=V_{\mathrm{mix}}^{N,N+2}=\omega\frac{\sqrt{(N+2)(N+1)}}{1+\beta^2}\Big(1+\frac{\sqrt{5}}{5}\beta^2\Big)\, .
\end{eqnarray}
After diagonalizing the matrix (\ref{VMatrix}), the resulting lowest eigenpotential, which comprises the lowest eigenvalue for each given $\beta$ and $\gamma$, is expected to correspond to the ground-state potential of the mixed system. Undoubtedly, the formulations provided in Eq.~(\ref{HCM})-Eq.~(\ref{Vmixpotential}) for the IBM-CM description of the 0p-0h and 2p-2h configurations can be directly extended to incorporate additional configurations, such as those arising from 4p-4h and 6p-6h excitations. In such a case, the total Hamiltonian and potential matrices would take on blocked forms of $3\times3$ and $4\times4$, respectively.

\begin{figure}
\begin{center}
\includegraphics[scale=0.15]{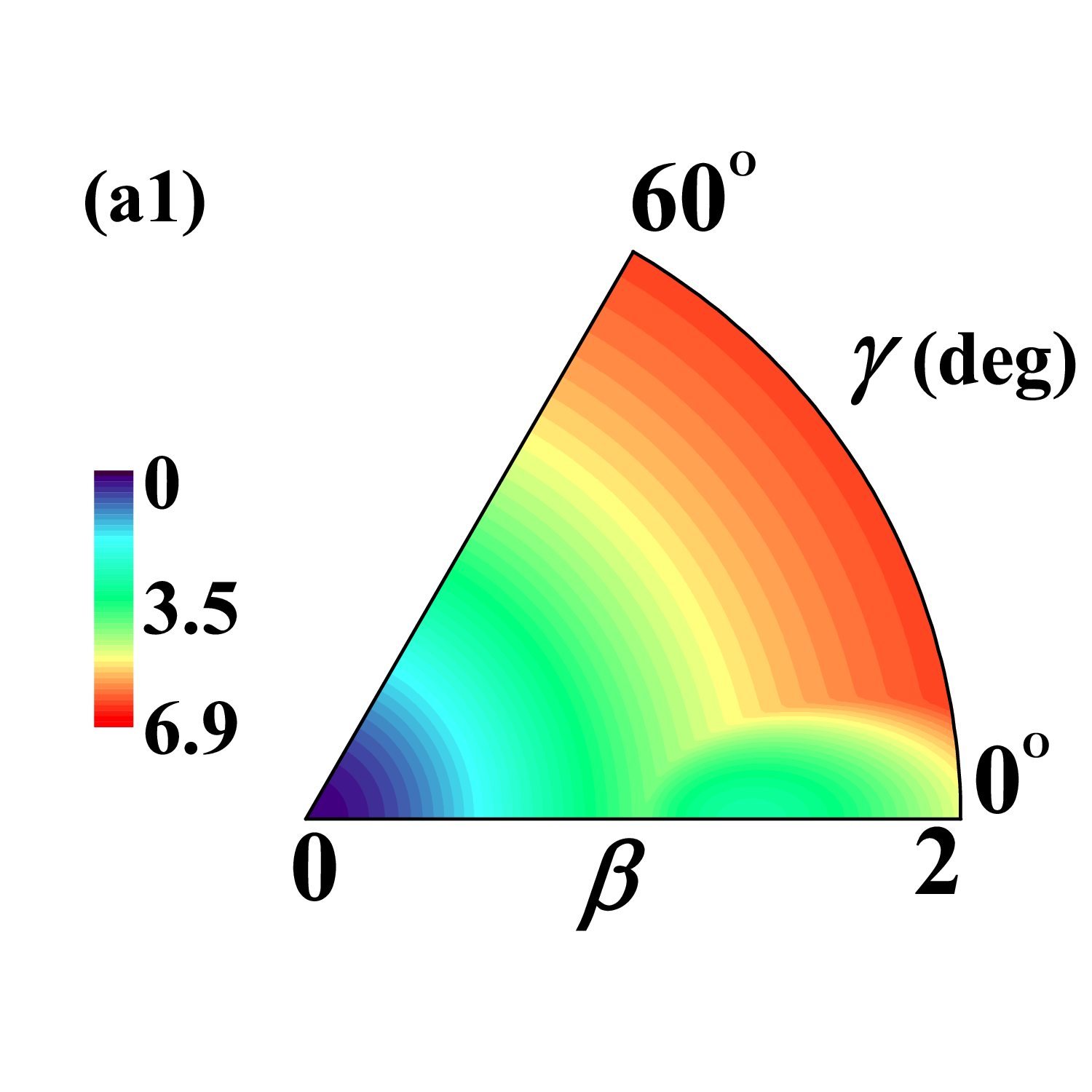}
\includegraphics[scale=0.15]{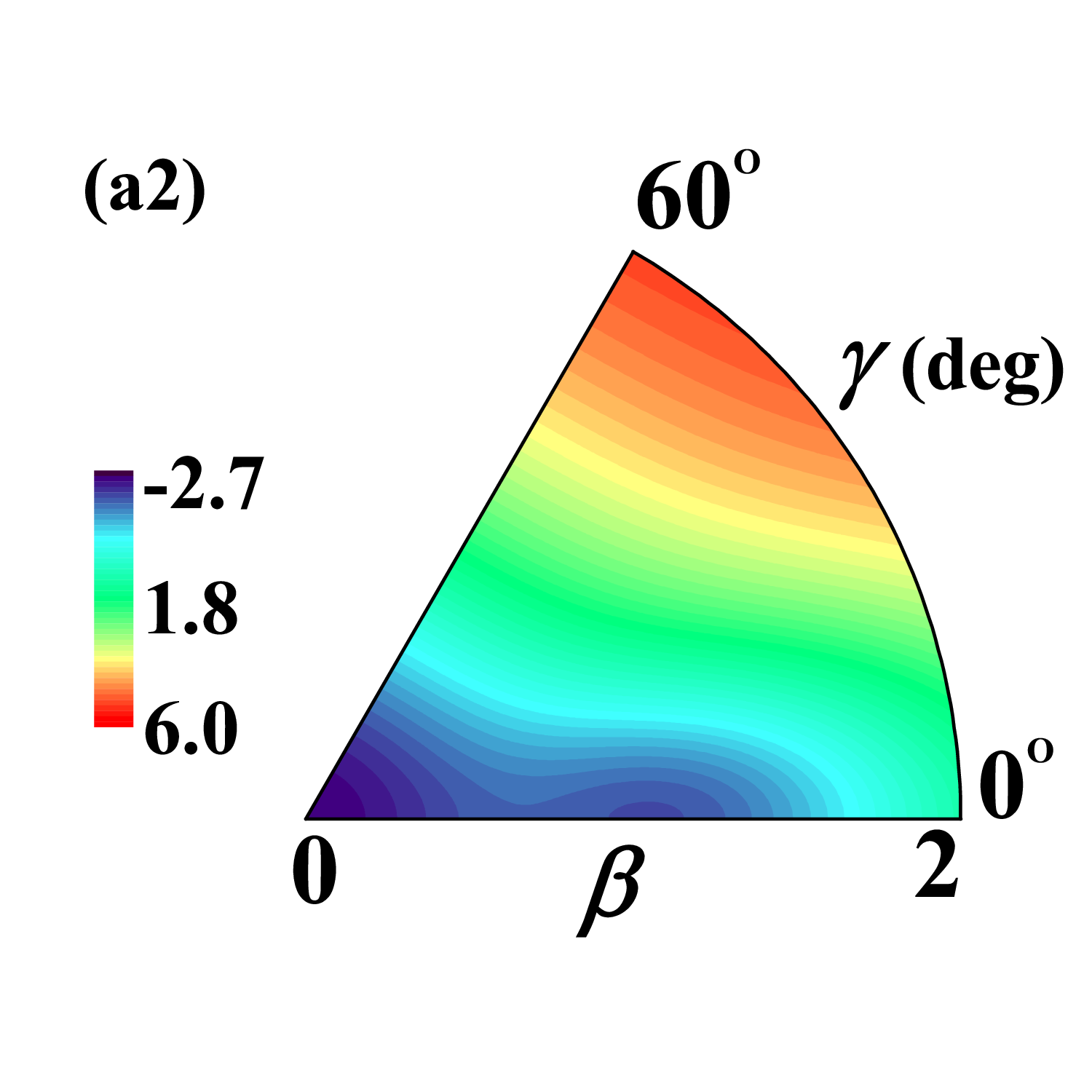}
\includegraphics[scale=0.15]{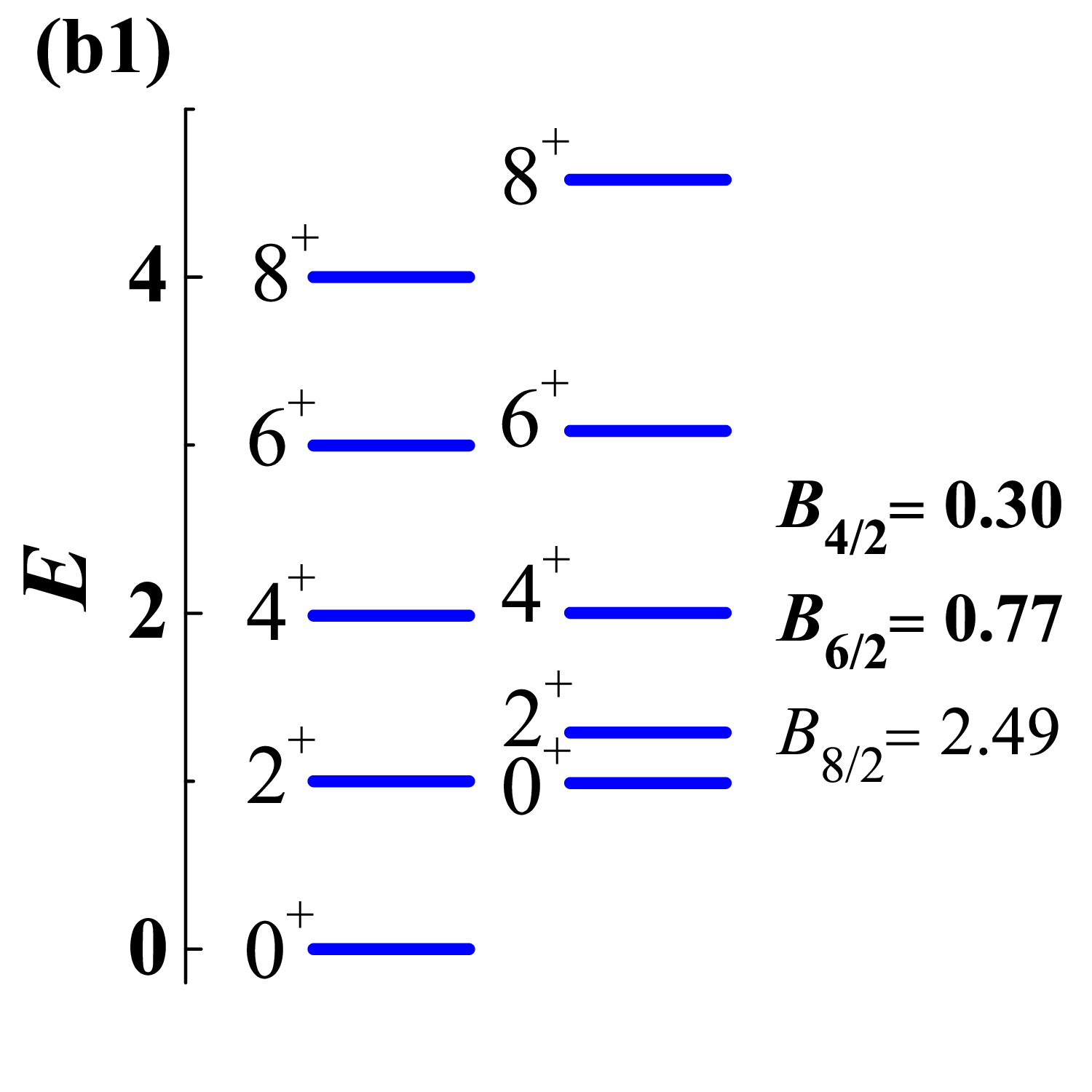}
\includegraphics[scale=0.15]{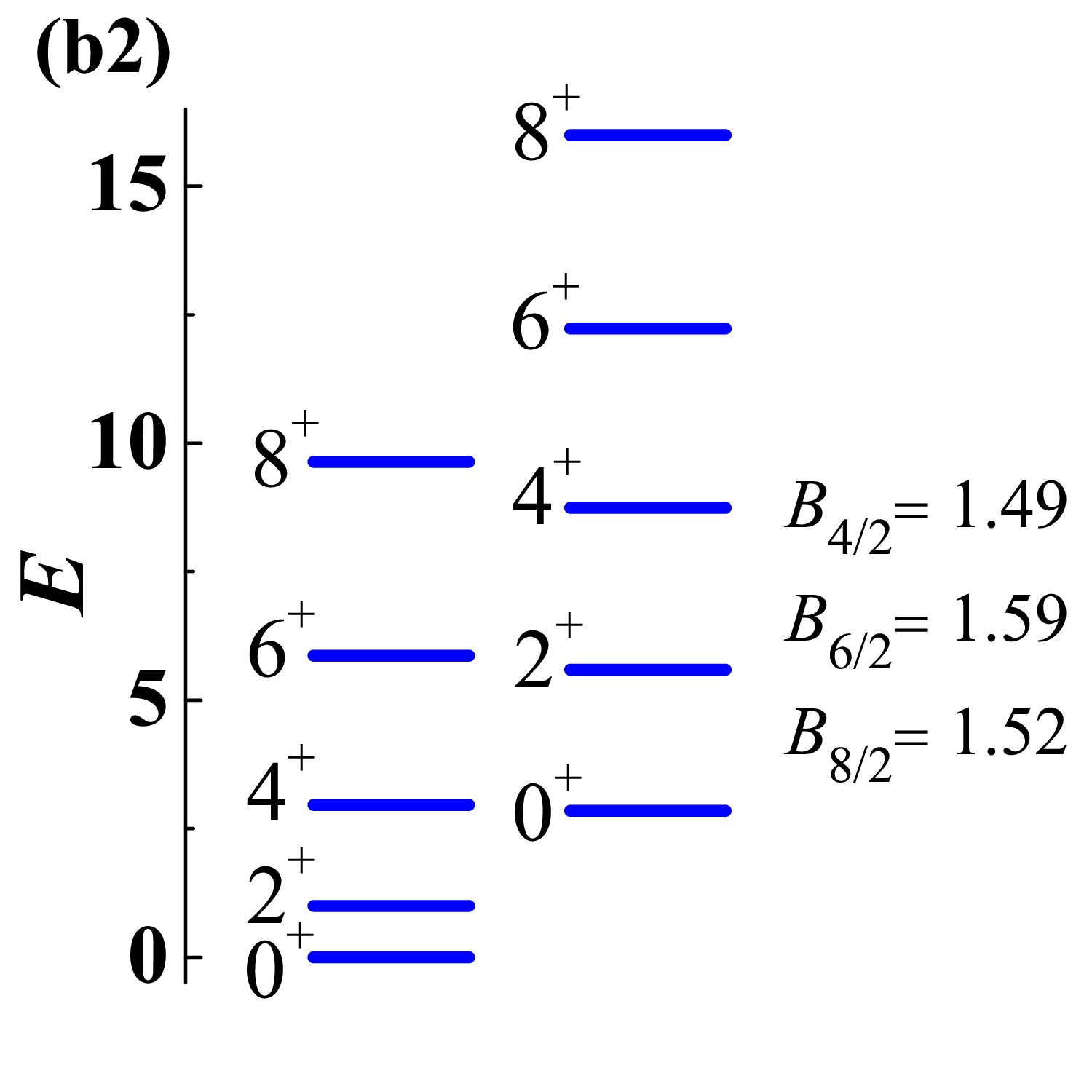}
\caption{Configuration mixing between U(5) and SU(3): (a1) The potential surface solved from (\ref{VMatrix}) with the nonzero parameters (in arbitrary units) set as $a_1^A=1.0,~a_2^B=-0.133,~\omega=0.006,~\Delta=31.58$, along with $(\chi^A,\chi^B)=(0,-\sqrt{7}/2)$, describing weak mixing between the normal (U(5)) configuration and intruder (SU(3)) configuration. (a2) The same as in (a1) but solved with $\omega=0.9$, describing strong mixing between the U(5)and SU(3) configurations. (b1) The level pattern (normalized to $E(2_1^+)=1.0$) corresponding to the case in (a1). (b2) The level pattern corresponding to the case (a2).}\label{F3}
\end{center}
\end{figure}

\begin{figure}
\begin{center}
\includegraphics[scale=0.15]{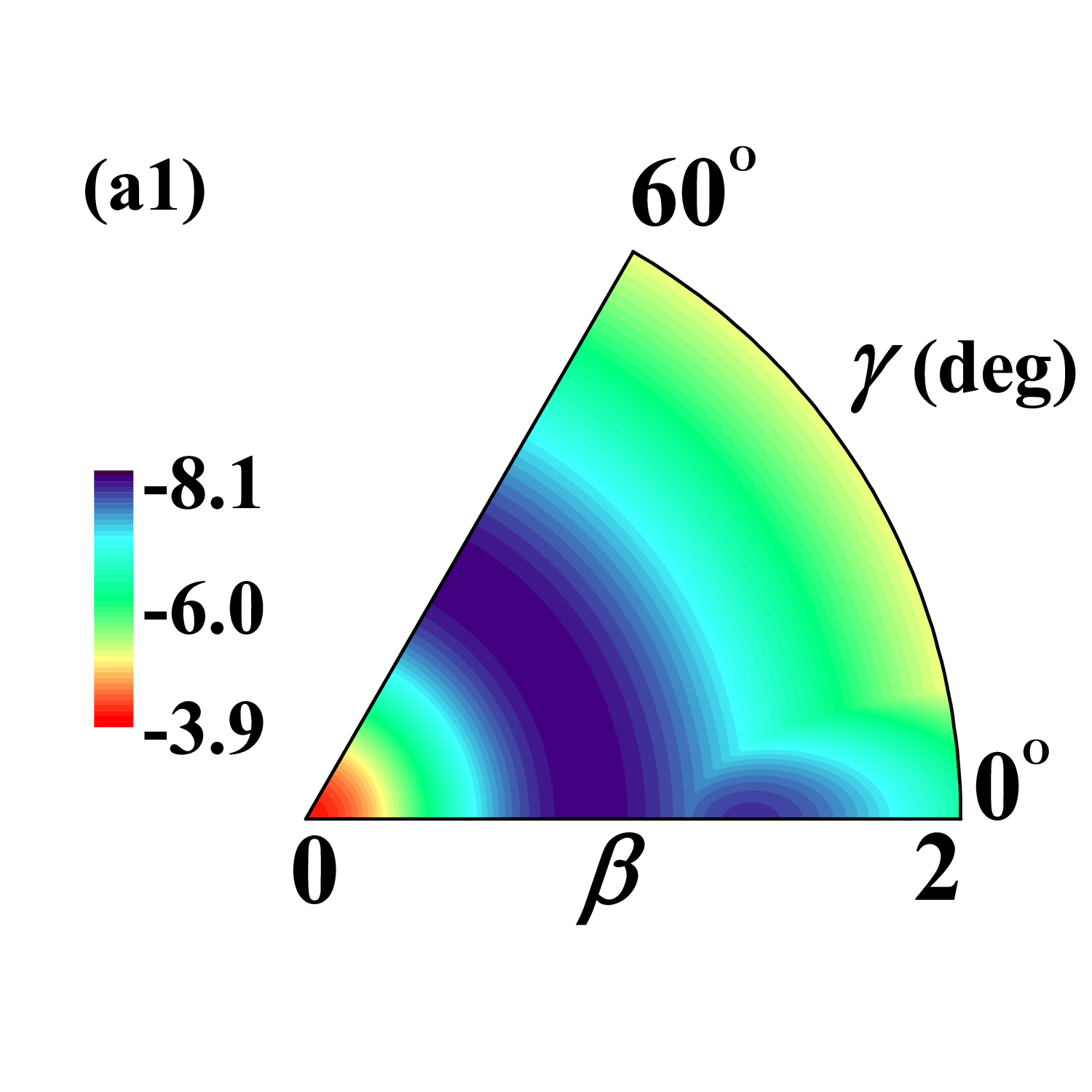}
\includegraphics[scale=0.15]{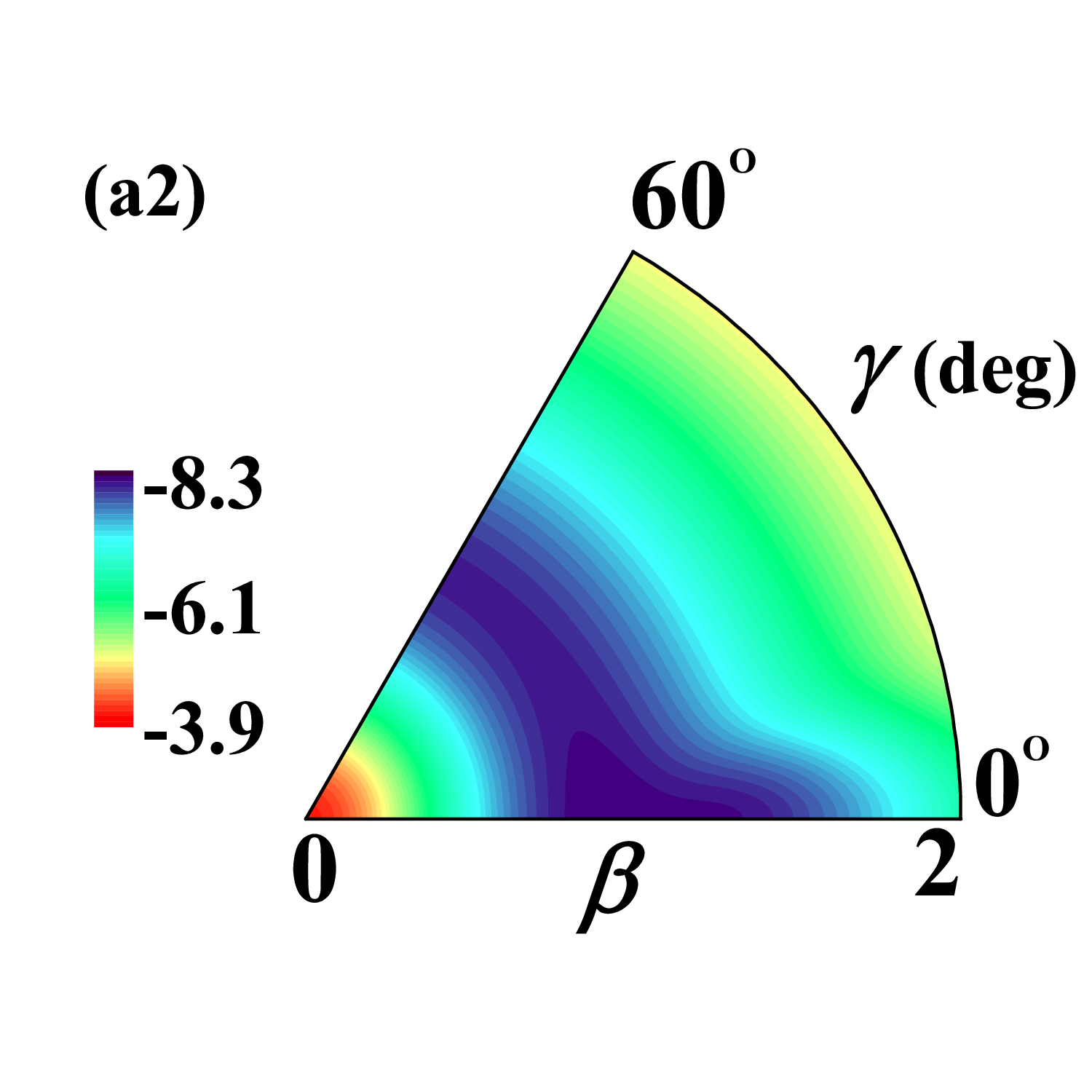}
\includegraphics[scale=0.15]{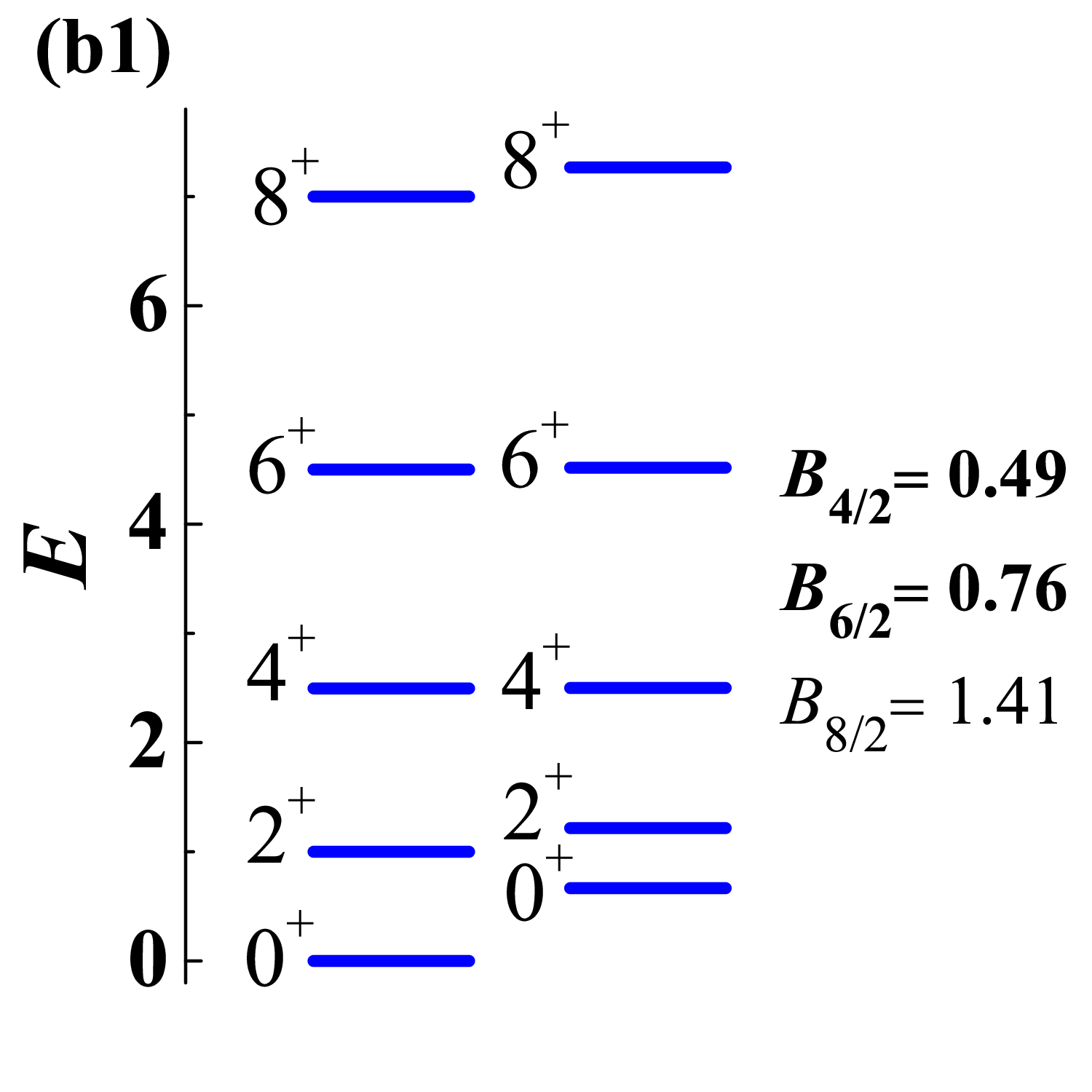}
\includegraphics[scale=0.15]{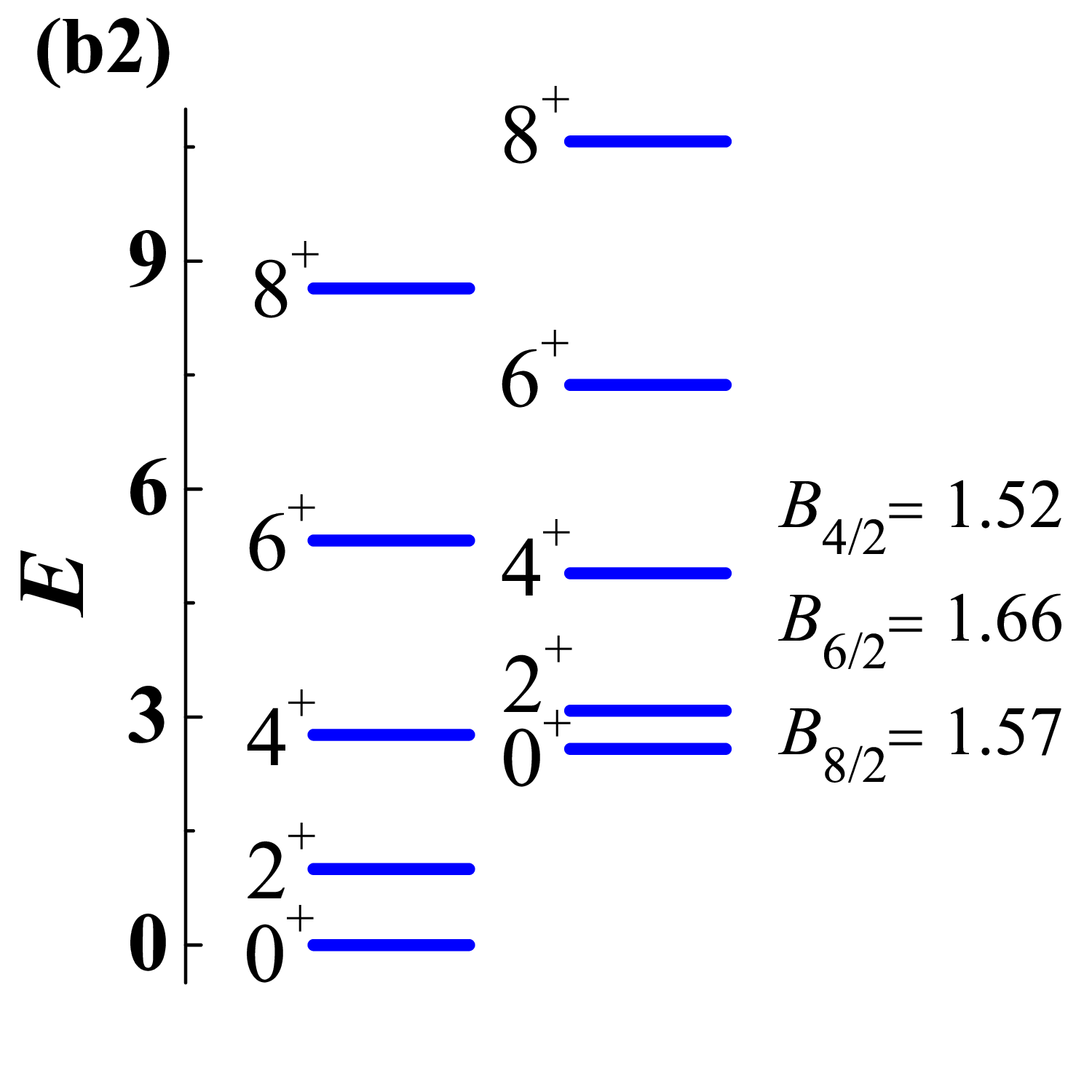}
\caption{Configuration mixing between O(6) and SU(3): (a1) The potential surface solved from (\ref{VMatrix}) with the nonzero parameters (in arbitrary units) set as $a_2^A=-0.1,~a_2^B=-0.0978,~\omega=0.0002,~\Delta=13.16$ along with $(\chi^A,\chi^B)=(0,-\sqrt{7}/2)$, describing weak mixing between the O(6) and SU(3) configurations. (a2) The same as in (a1) but solved with $\omega=0.1$, describing strong mixing between the O(6)and SU(3) configurations. (b1) The level pattern (normalized to $E(2_1^+)=1.0$) corresponding to the case in (a1). (b2) The level pattern corresponding to the case in (a2).}\label{F4}
\end{center}
\end{figure}

\begin{figure}
\begin{center}
\includegraphics[scale=0.15]{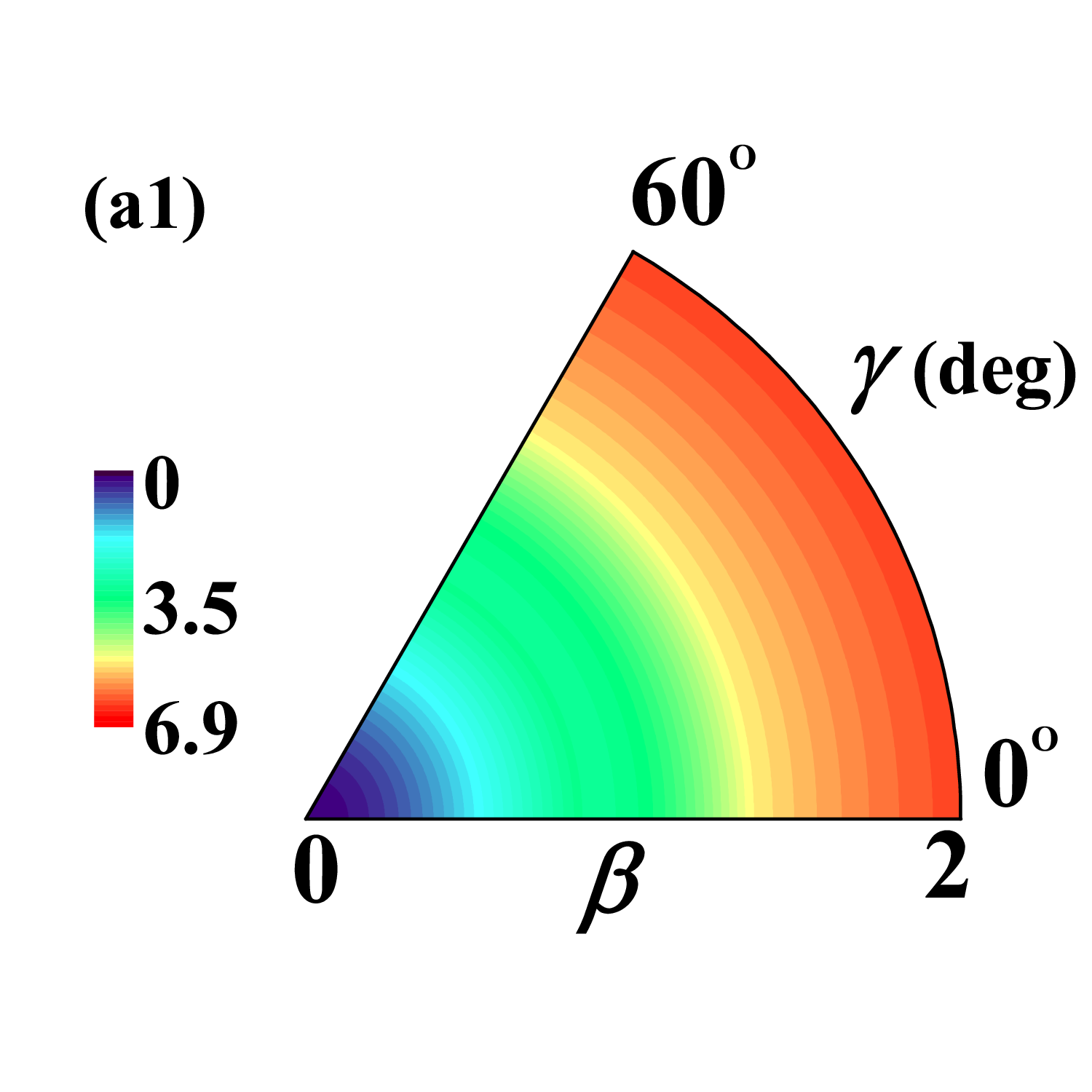}
\includegraphics[scale=0.15]{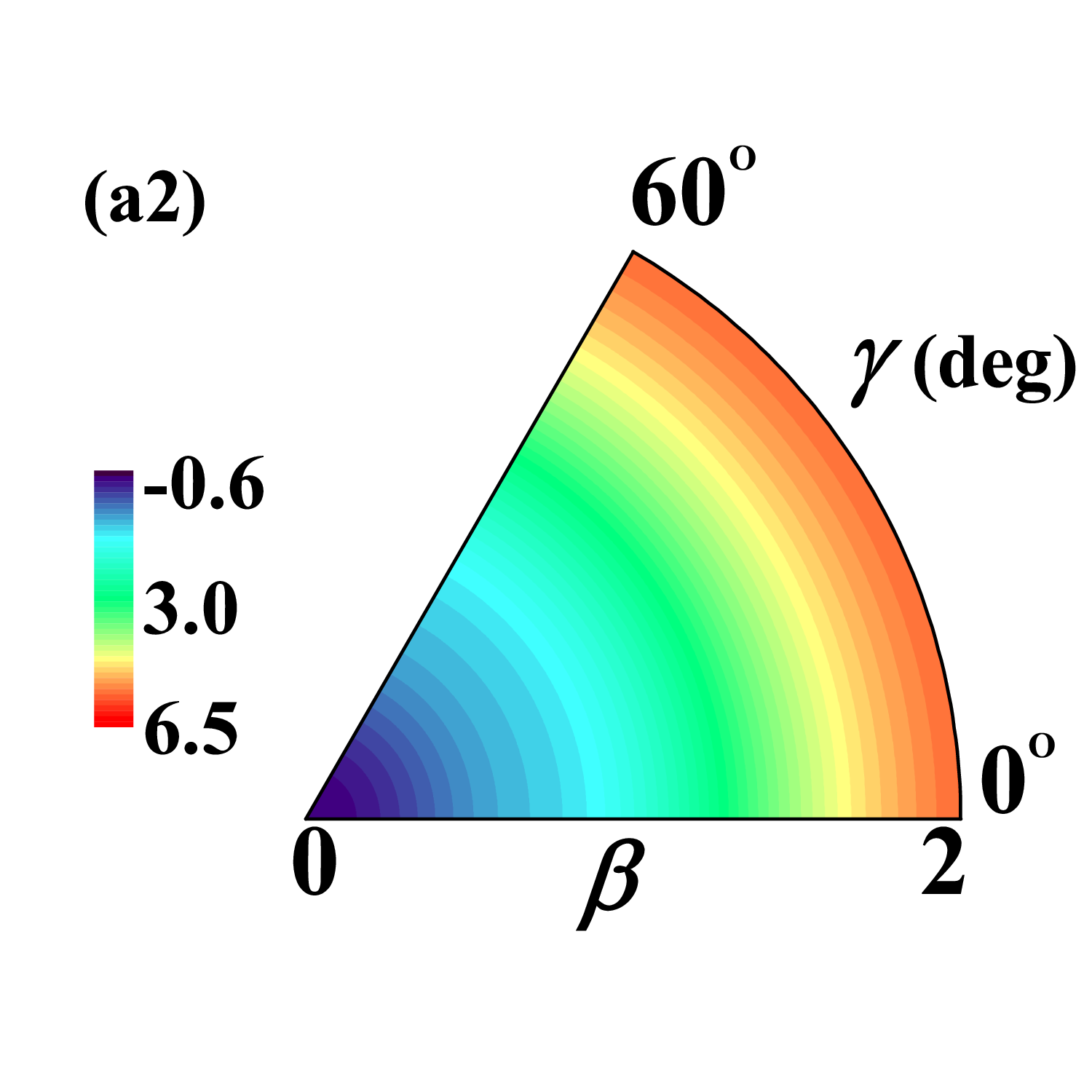}
\includegraphics[scale=0.15]{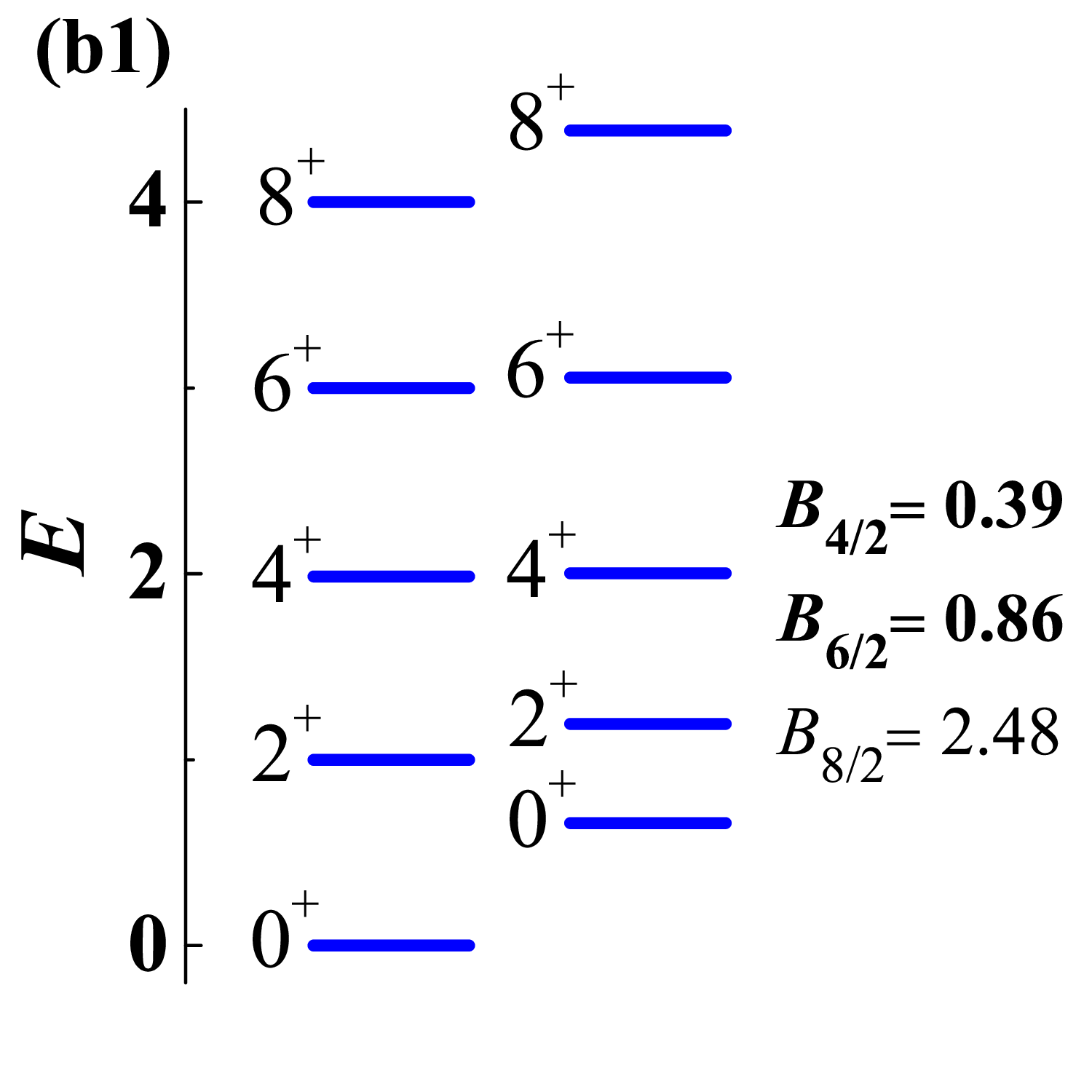}
\includegraphics[scale=0.15]{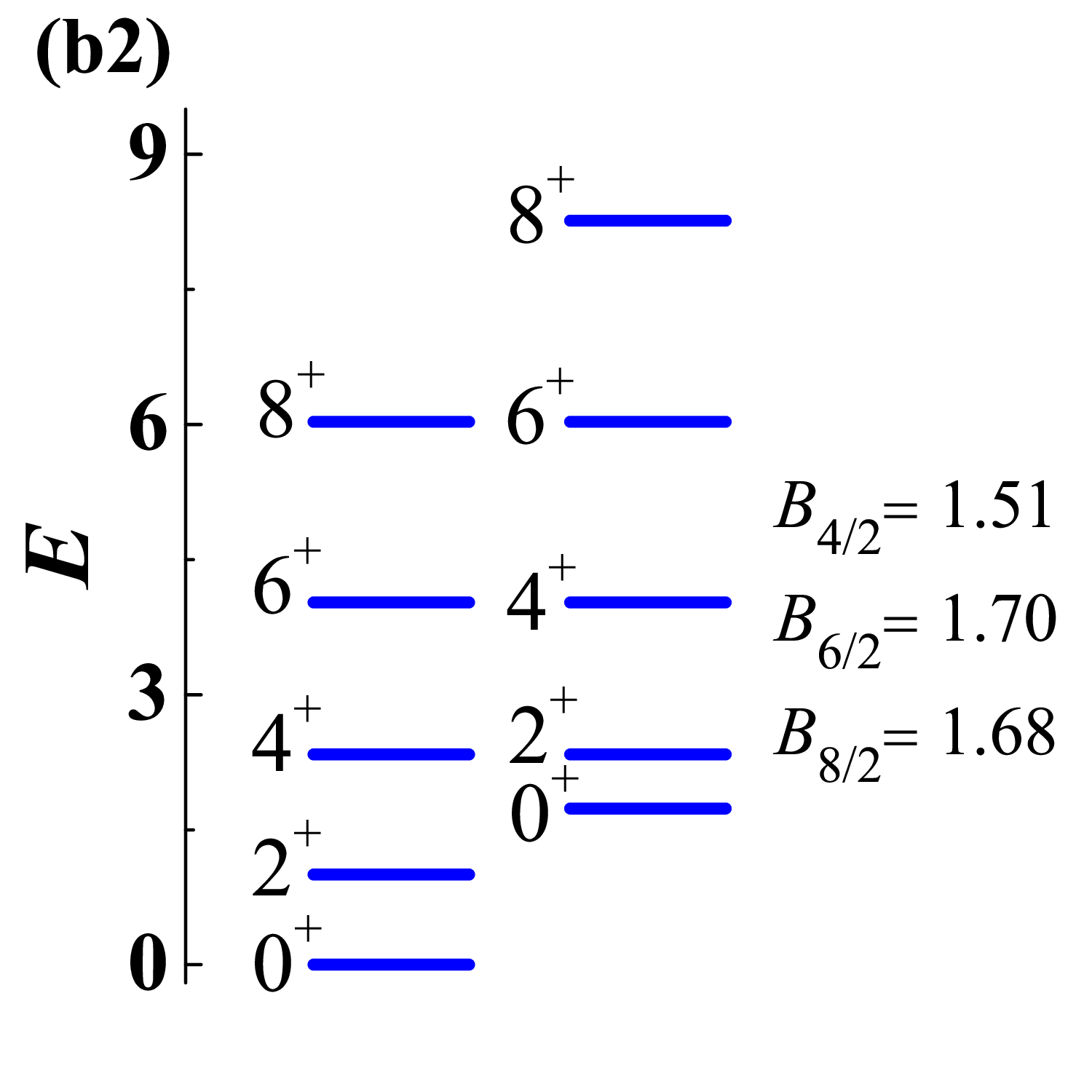}
\caption{Configuration mixing between U(5) and O(6): (a1) The potential surface solved from (\ref{VMatrix}) with the nonzero parameters (in arbitrary units) set as $a_1^A=1.0,~a_2^B=-0.133,~\omega=0.0015,~\Delta=19.28$ along with $(\chi^A,\chi^B)=(0,0)$, describing the weak mixing between the U(5) and O(6) configurations. (a2) The same as in (a1) but solved with $\omega=0.3$, describing strong mixing between the U(5) and O(6) configurations. (b1) The level pattern (normalized to $E(2_1^+)=1.0$) corresponding to the case in (a1). (b2) The level pattern corresponding to the case in (a2).}\label{F5}
\end{center}
\end{figure}

The previous IBM-based analyses~\cite{Zhang2022,Pan2024,Teng2025,Zhang2025} demonstrate that $B(E2)$ anomaly in yrast states may arise from a triaxial system with strong band mixing, which can be modeled by introducing the three-body $LQL$ and four-body $LQQL$ terms into the Hamiltonian. If intruder configuration is favored at low energy, as indicated by the IBM-CM descriptions, the intruder states may be pulled down to affect the $B(E2)$ structures of the yrast band. To distinguish this scenario from the band-mixing cases without configuration mixing discussed above, we focus below on analyzing how a $B_{4/2}<1.0$ result can be produced from the mixing between the two "ground-state" bands built on $0_1^+$ and $0_2^+$ in the IBM-CM scheme, where $0_2^+$ corresponds to the "ground state" of the intruder configuration in the zero mixing ($\omega=0$) limit. In the following, we take the boson system with $N=8$ corresponding to $^{172}$Pt as an example for theoretical discussions. First, we examine the two-band patterns generated by the IBM-CM calculations using the parameters determined in early studies~\cite{Harder1997,Morales2008,Ramos2009} conducted prior to the revelation of the anomalous result $B_{4/2}\simeq0.55(19)$ in $^{172}$Pt~\cite{Cederwall2018}. These analysis~\cite{Harder1997,Morales2008,Ramos2009,Ramos2011,Ramos2014} indicate that the spectral properties of both light and heavy Pt isotopes can be well captured within the IBM-CM framework incorporating the 0p-0h and 2p-2h configurations. Specifically, two sets of parameters (in MeV) are adopted here to model the $^{172}$Pt system: set I, taken from Refs. \cite{Harder1997,Morales2008}, is given by $(a_1^A,a_1^B)=(0.54,0)$, $(a_2^A,a_2^B)=(-0.027,-0.022)$ and $(b_2^A,b_2^B)=(0,0.01)$, along with the dimensionless parameters $(\chi^A,\chi^B)=(0.25,-0.45)$; set II, taken from \cite{Ramos2009}, is given by $(a_1^A,a_1^B)=(0.725,0)$, $(a_2^A,a_2^B)=(-0.03947,-0.02287)$, $(b_2^A,b_2^B)=(0,0)$, together with $(\chi^A,\chi^B)=(0,-0.38)$. In both cases, the parameters for the mixing term and the off-set energy are set to $\omega=0.05$ MeV and $\Delta=2.8$ MeV. Note that the value of $\Delta=1.4$ MeV quoted in \cite{Harder1997} corresponds to half the off-set energy, as pointed out in \cite{Morales2008}. Since the cubic term and three-body rotor-like term $LQL$ were not included in the earlier IBM-CM schemes~\cite{Harder1997,Morales2008,Ramos2009}, the corresponding parameters will constantly set to $a_3^A=a_3^B=b_1^A=b_1^B=0$ unless specifically noted. With this parametrization, the resulting mean-field potentials and spectral patterns, including the ratios $B_{L/2}\equiv B(E2;L_1^+\rightarrow(L-2)_1^+)/B(E2;2_1^+\rightarrow0_1^+)$, are presented in Fig.~\ref{F2}. To demonstrate the $B_{4/2}<1.0$ feature obtained from the IBM-CM calculations, results for two additional cases derived by further adjusting the parameters based on case II (parameters in set II) are also provided in Fig.~\ref{F2}. Specially, case III is achieved by reducing the off-set energy to $\Delta=2.04$ MeV and setting $\omega=0$, corresponding to a zero-mixing between the two configurations, while case IV is obtained by resetting the parameters with $\Delta=1.718$ MeV, $\omega=0.005$ MeV and $b_2^B=0.016$ MeV, which describes weak configuration mixing. For calculating the $B(E2)$ transitions, the effective charges in Eq.~(\ref{E2CM}) have been set to $e_A=e_B$ in all cases for simplicity. It should be emphasized again that our focus here is on discussing the possibility of $B(E2)$ anomaly solely arising from the IBM-CM rather than finding reasonable parameters for describing experimental data.

As observed in Fig.~\ref{F2}, the mean-field landscapes derived from different parameter sets are globally similar, all displaying U(5)-like potentials. This is actually consistent with the large $a_1^A$ values adopted in all cases. Note that the potential surface shown in Fig.~\ref{F2}(I1) was also derived in the previous study~\cite{Ramos2014}, where a detailed investigation of the potential surfaces and nuclear shapes in Pt isotopes was carried out within the IBM-CM framework. As indicated in Fig.~\ref{F2}(III1) and Fig.~\ref{F2}(IV1), a deformed local minimum may appear at low energy when adopting smaller $\Delta$ value. This, in turn, results in a significantly reduced $B_{4/2}$ value, consistent with the observed $B_{4/2}=0.55(19)$ in $^{172}$Pt~\cite{Cederwall2018}, as demonstrated by the results presented in Fig.~\ref{F2}(IV2). Such an outcome is achieved by adjusting the off-set energy parameter to lower the first $4^+$ level in configuration $B$ to the yrast sequence, as shown in Fig.~\ref{F2}(III2). In this case, the yrast states above $E(2_1^+)$ originate from the intruder configuration, while $0_1^+$ and $2_1^+$ belong to the normal configuration, leading to $B_{4/2}=0$ due to level crossing. Clearly, such a $B(E2)$ anomaly feature arises in a situation with very weak configuration mixing, which is caused by a suitable choice of the $\Delta$ value. Moreover, the suppressed $B(E2)$ transitions are established only for specific $L$ values and remain significantly stronger for others, as illustrated in Fig.~\ref{F2}(IV2). In contrast, normal $B(E2)$ structures with $B_{L/2}>1.0$ arise in cases of intermediate or strong configuration mixing, as demonstrated in Fig.~\ref{F2}(I2) and Fig.~\ref{F2}(II2), where the yrare states $0_2^+$ and $2_2^+$ can be regarded as the band heads of the quasi-$\beta$ and quasi-$\gamma$ bands, respectively. These results are consistent with prior calculations performed under the same parameters~\cite{Harder1997,Ramos2009}.

To better understand the $B(E2)$ anomaly produced in the IBM-CM schemes under weak mixing, we provide below several examples to illustrate configuration mixing between the symmetry limits of the IBM. Accordingly, the Hamiltonian is constructed by assuming that the normal configuration $A$ is described by either the U(5) or O(6) limits, and the intruder configuration $B$ is described by either the SU(3) or O(6) limits. For comparison, the potential surfaces and level patterns solved under both weak and strong mixing conditions are shown for each case. As seen in Fig.~\ref{F3}(b1), significant suppressed $B(E2)$ values with $B_{4/2}<1.0$ and $B_{6/2}<1.0$ indeed appear in the configuration mixing between U(5) and SU(3) limits, which simultaneously yields a large $B_{8/2}$ ratio value. Undoubtedly, this unusual $B(E2)$ feature primarily arises from level crossing between two configurations, which is achieved through an appropriate choice of the off-set energy $\Delta$. Meanwhile, this parametrization may lead to the approximate degeneracies between yrast and yrare levels for a given $L$, including $E(4_1^+)\approx E(4_2^+)$ and $E(6_1^+)\approx E(6_2^+)$. On the other hand, these predicted $B(E2)$ anomaly feature and the associated approximate degeneracies vanish in the strong mixing case, as shown in Fig.~\ref{F3}(b2). Instead, a spherical-prolate shape coexistence emerges, as evidenced by the potential surface presented in Fig.~\ref{F3}(a2). In contrast, the potential in the weak mixing case exhibits two isolate minima located at different energies, implying well-separated U(5) and SU(3) modes.

Situations similar to those depicted in Fig.~\ref{F3} can also be observed in Fig.~\ref{F4} and Fig.~\ref{F5}, where the $B(E2)$ anomaly features arise from weak mixing between O(6) and SU(3), or between U(5) and O(6). Particularly, the results in Fig.~\ref{F5} suggest that the presence of $B(E2)$ anomaly is not easy to be discerned from observing the mean-field potential when the two configurations have relatively close dynamic components, such as both the U(5) and O(6) limits sharing the common O(5) subgroup symmetry and thus being $\gamma$-unstable at the mean-field level~\cite{Iachellobook}.
However, the universal $B_{L/2}<1.0$ feature along the yrast line, as predicted by the previous IBM analyses~\cite{Zhang2022,Wang2020,Zhang2024,Pan2024,Teng2025,Zhang2025}, cannot be fully reproduced solely through these IBM-CM schemes. This limitation can potentially be mitigated to some extent by incorporating rotor-like terms in the Hamiltonian. For strong configuration-mixing, it is not difficult to understand this point, as a system in such cases may behavior more like a single configuration and is therefore expected to be influenced by the rotor-like terms in a manner similar to cases without configuration mixing. For that, we briefly discuss an example based on the strong configuration-mixing between U(5) and SU(3) shown in Fig.~\ref{F3}. By adopting the same parameters as those used in Fig.~\ref{F3}(b2), except for additionally setting $b_1^A=b_1^B=-0.02$ and $b_2^B=0.303$ to account for contributions from the rotor-like terms, the calculation yields $B_{4/2}=0.78$, $B_{6/2}=0.72$, and $B_{8/2}=0.74$. This demonstrates that more $B(E2)$ anomaly values can indeed be obtained under conditions of strong mixing between two distinct configurations, provided that the rotor-like terms are included. However, if the rotor-like terms are directly added to the weak-mixing schemes described above, the parameter conditions required to sustain level crossing-and thereby maintain the $B_{L/2}<1.0$ feature-will be disrupted. The question then rises as to whether more $B(E2)$ anomaly results can be achieved under alternative weak configuration mixing scenarios. To explore this, we examine the following two possibilities: the inclusion of additional configurations and the combination of specific type of configuration-mixing with the rotor-like terms.

\begin{figure}
\begin{center}
\includegraphics[scale=0.15]{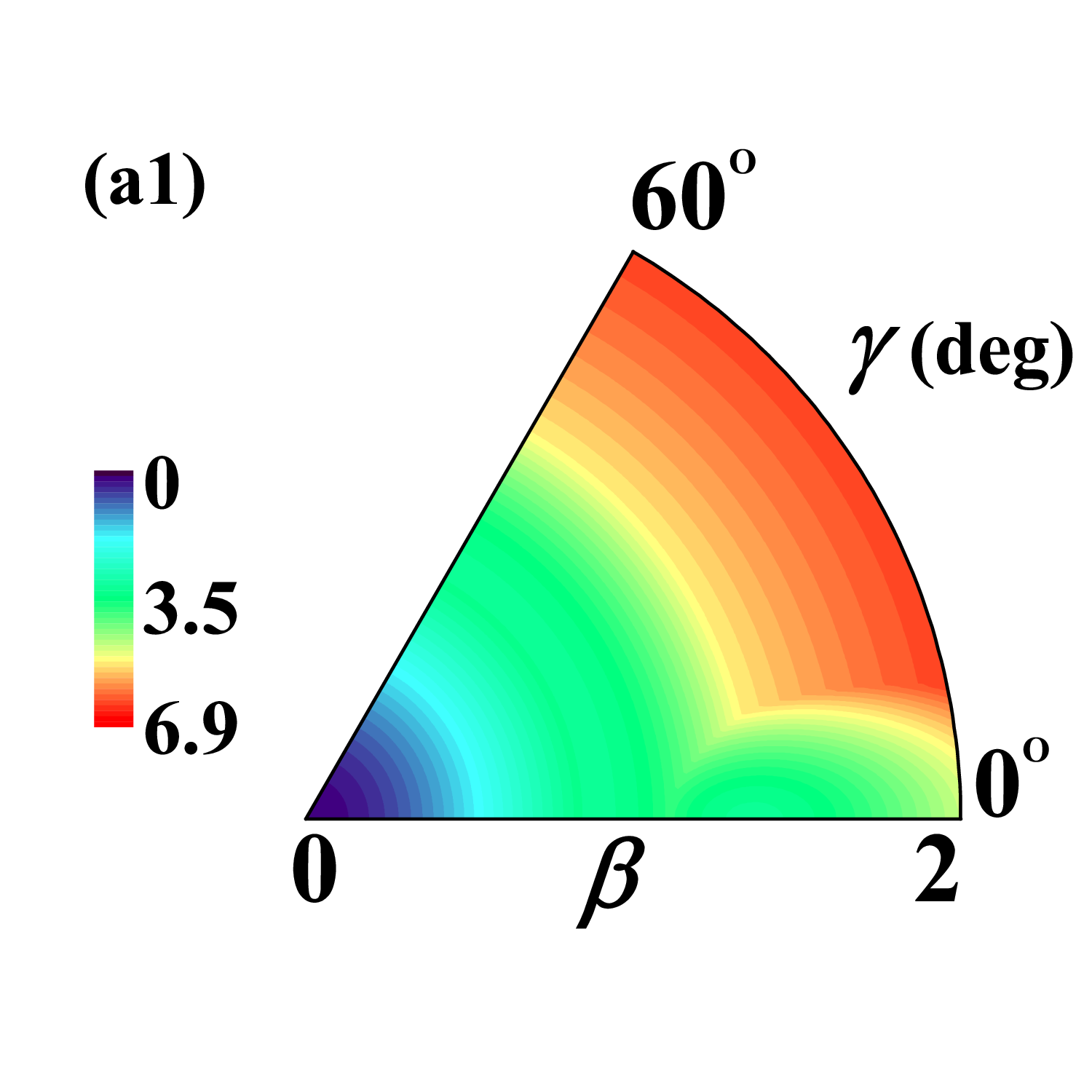}
\includegraphics[scale=0.15]{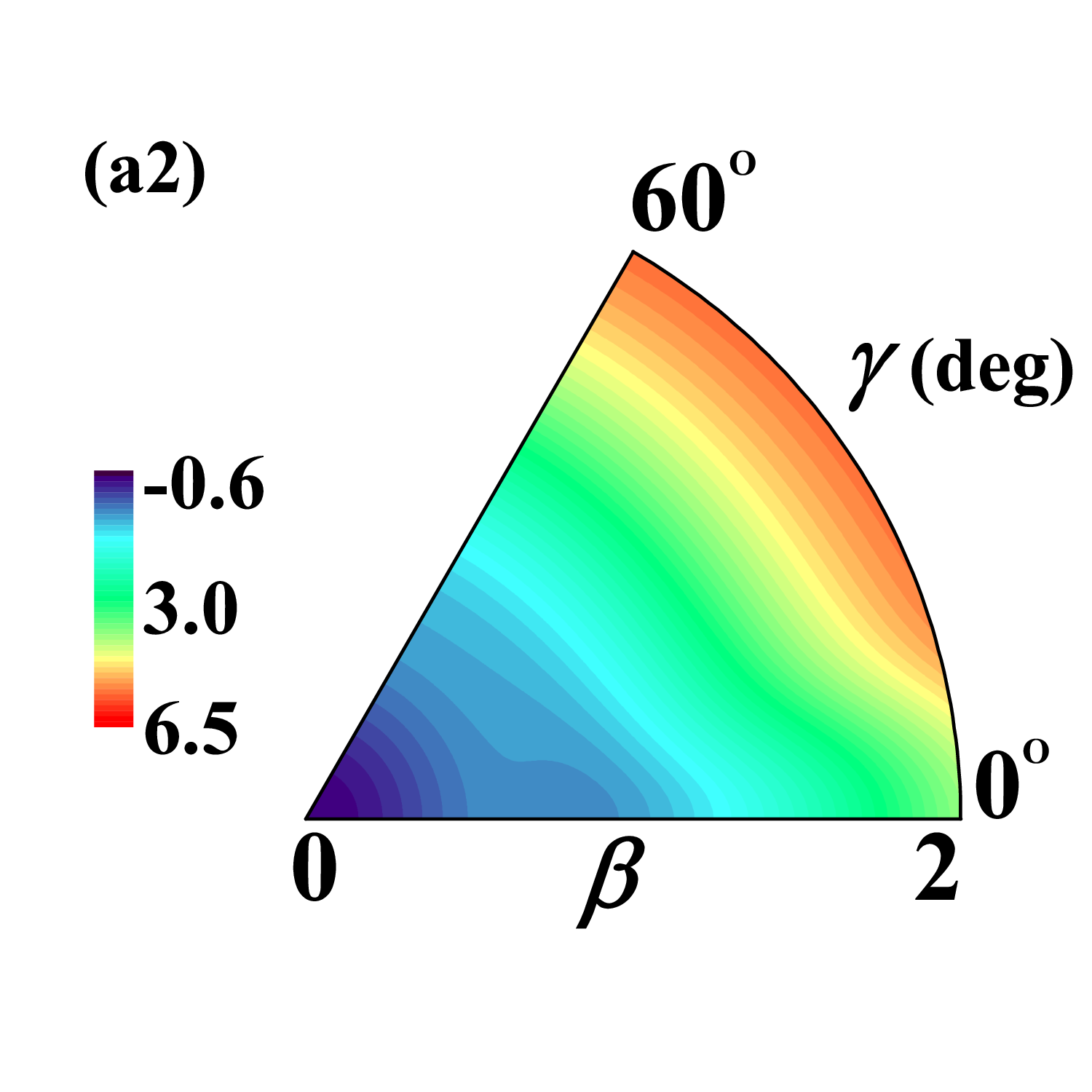}
\includegraphics[scale=0.15]{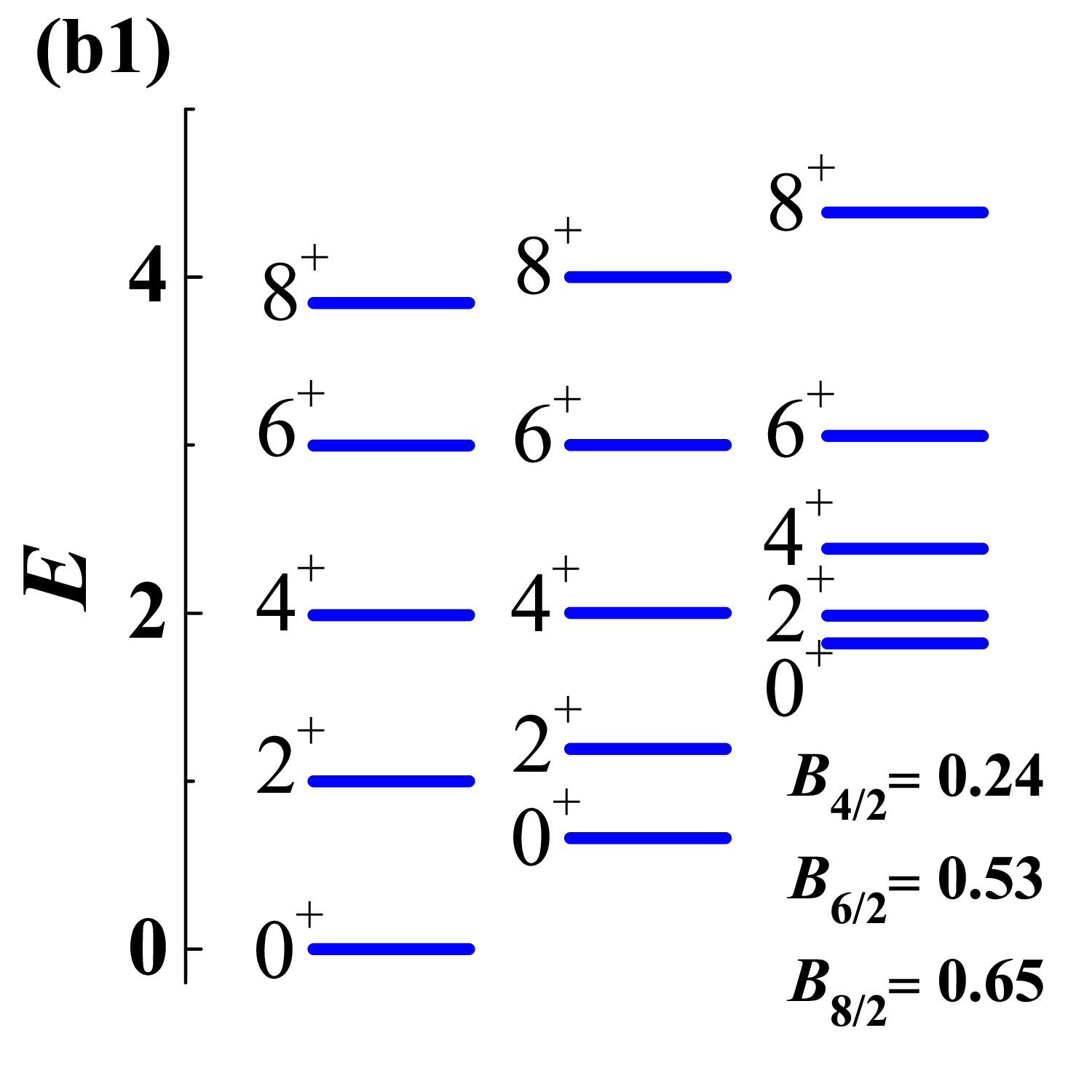}
\includegraphics[scale=0.15]{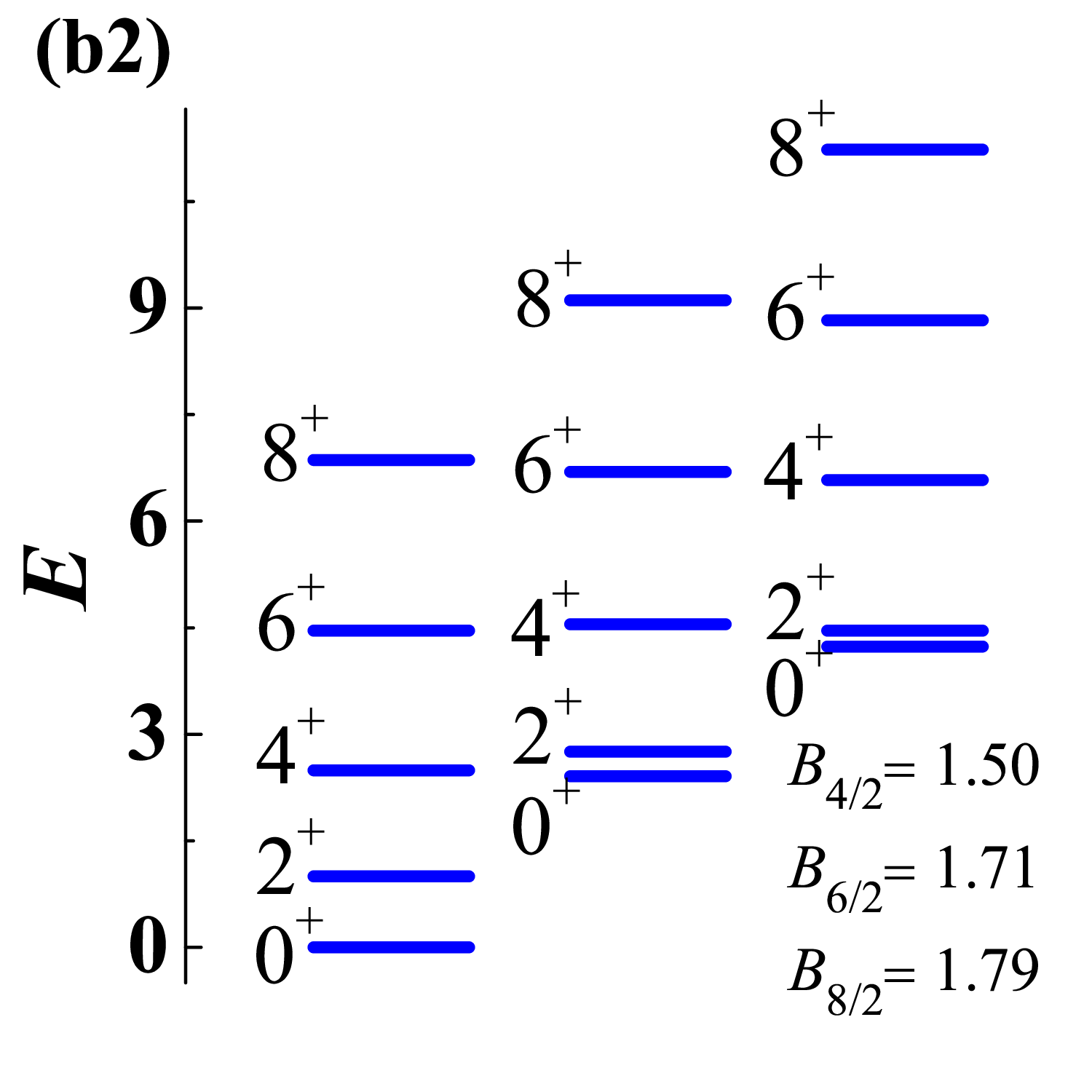}
\caption{Configuration mixing among U(5), O(6) and SU(3): (a1) The potential surface solved from (\ref{VMatrix}) in the weak mixing case. (a2) The same as in (a1) for the strong mixing case. (b1) The level pattern (normalized to $E(2_1^+)=1.0$) corresponding to the case in (a1). (b2) The level pattern corresponding to the case in (a2). The adopted parameters in each panel are detailed in the text. }\label{F6}
\end{center}
\end{figure}

\begin{figure*}
\begin{center}
\includegraphics[scale=0.16]{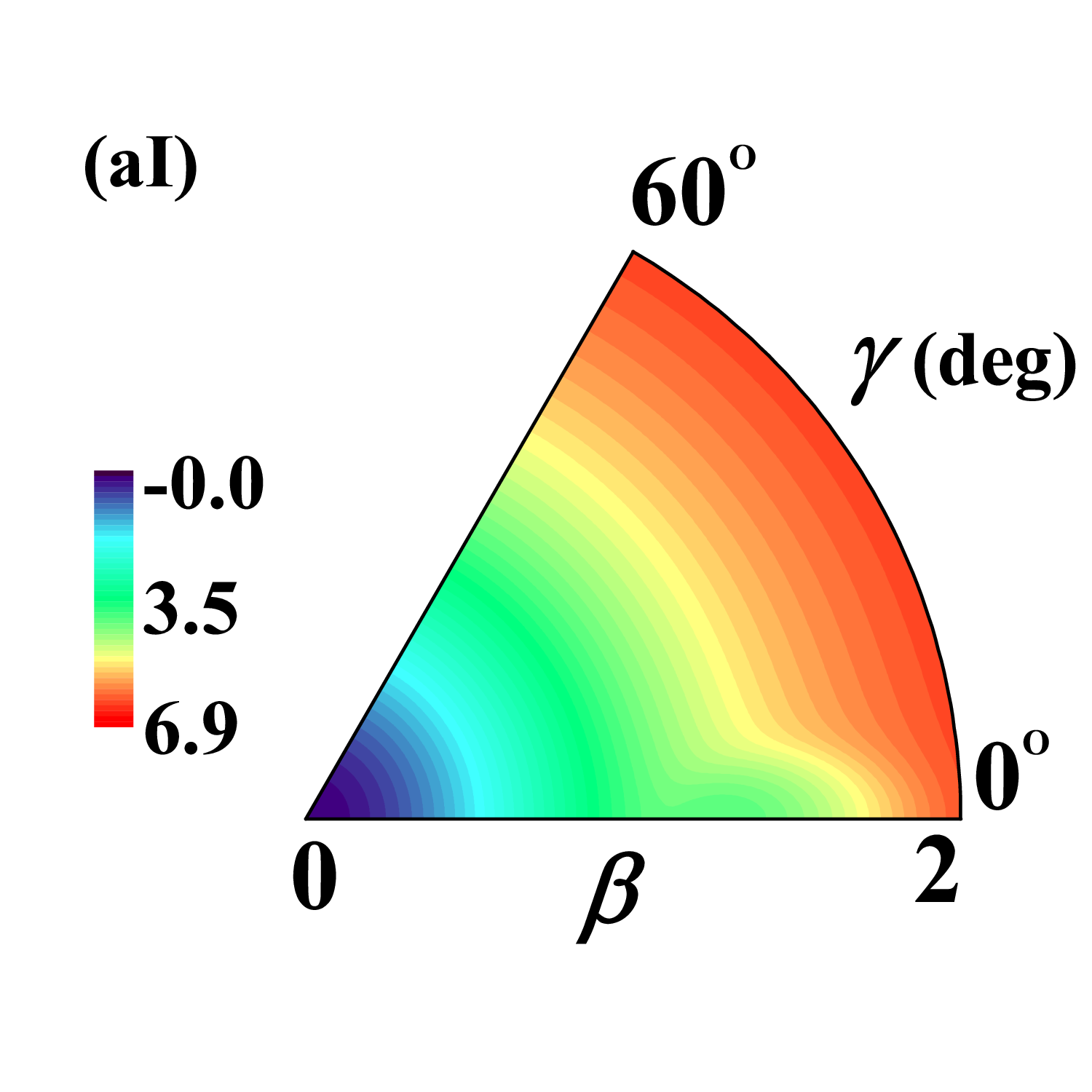}
\includegraphics[scale=0.16]{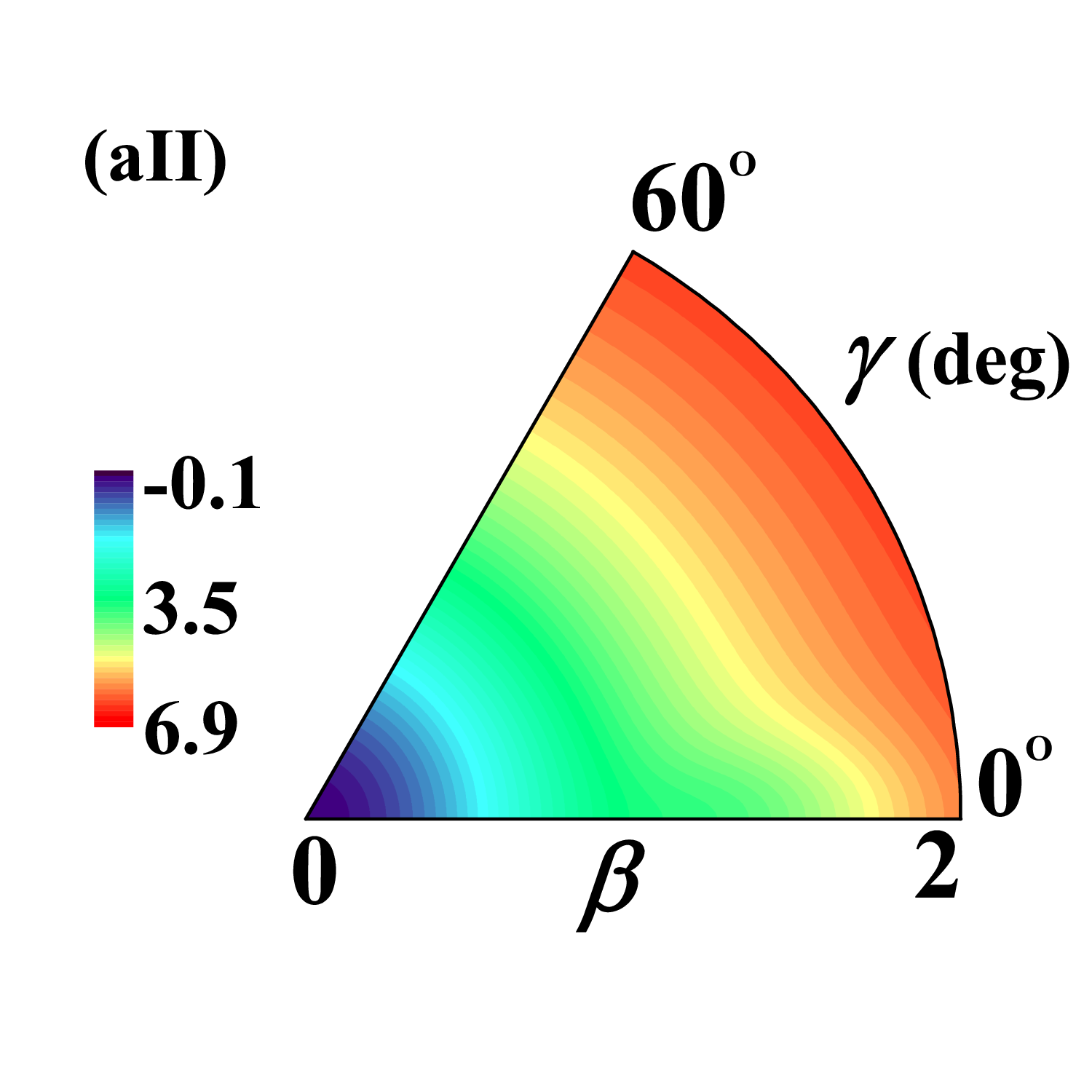}
\includegraphics[scale=0.16]{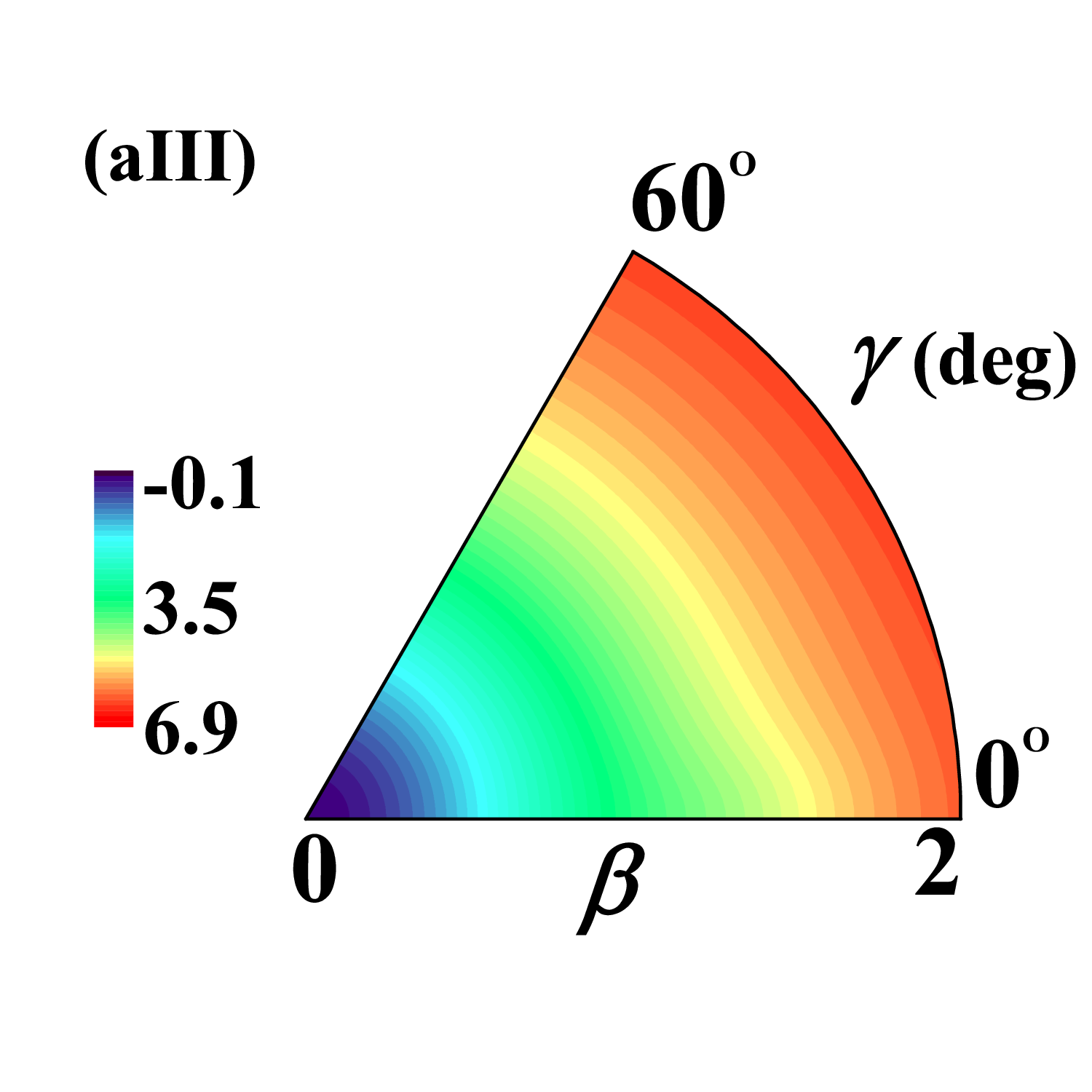}
\includegraphics[scale=0.16]{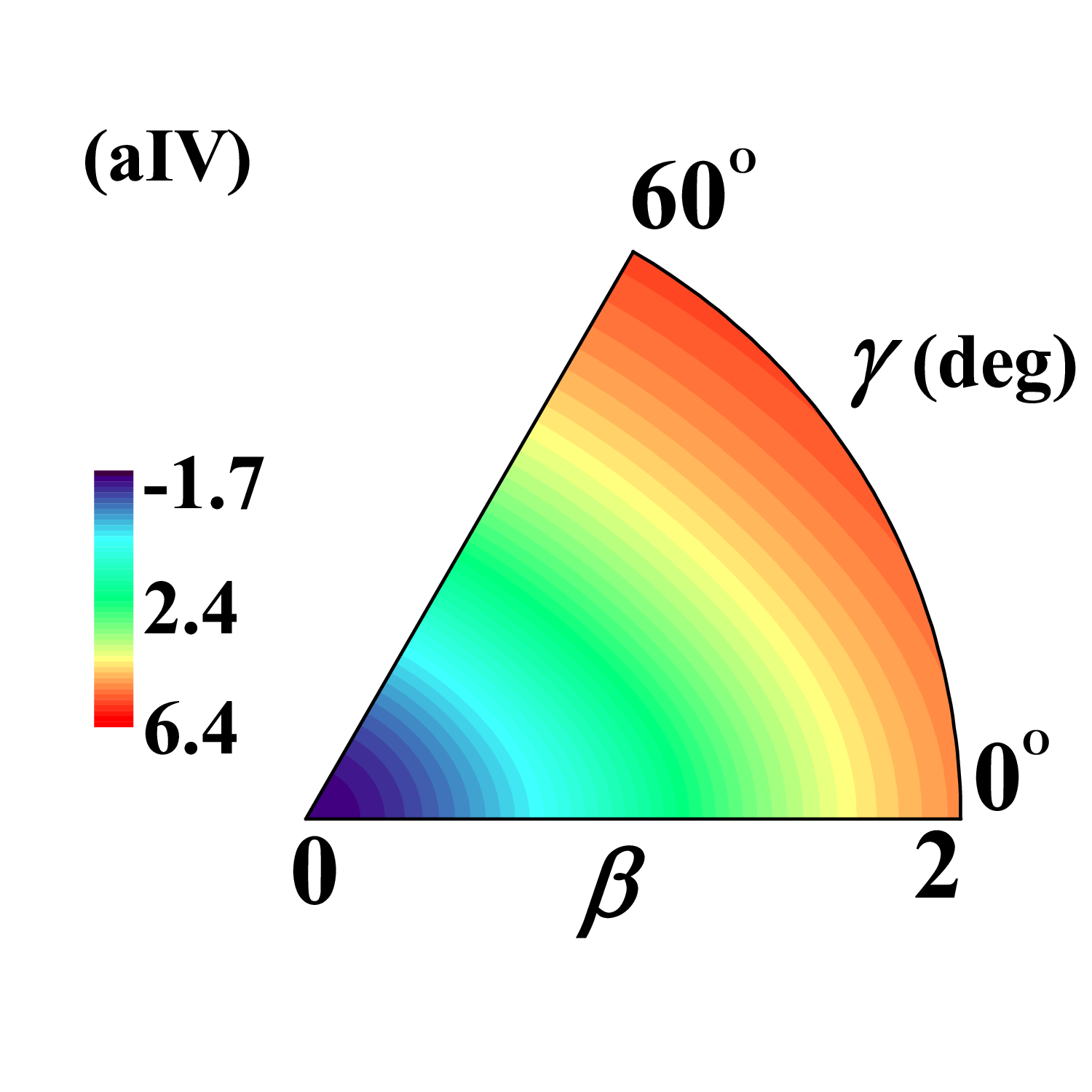}
\includegraphics[scale=0.16]{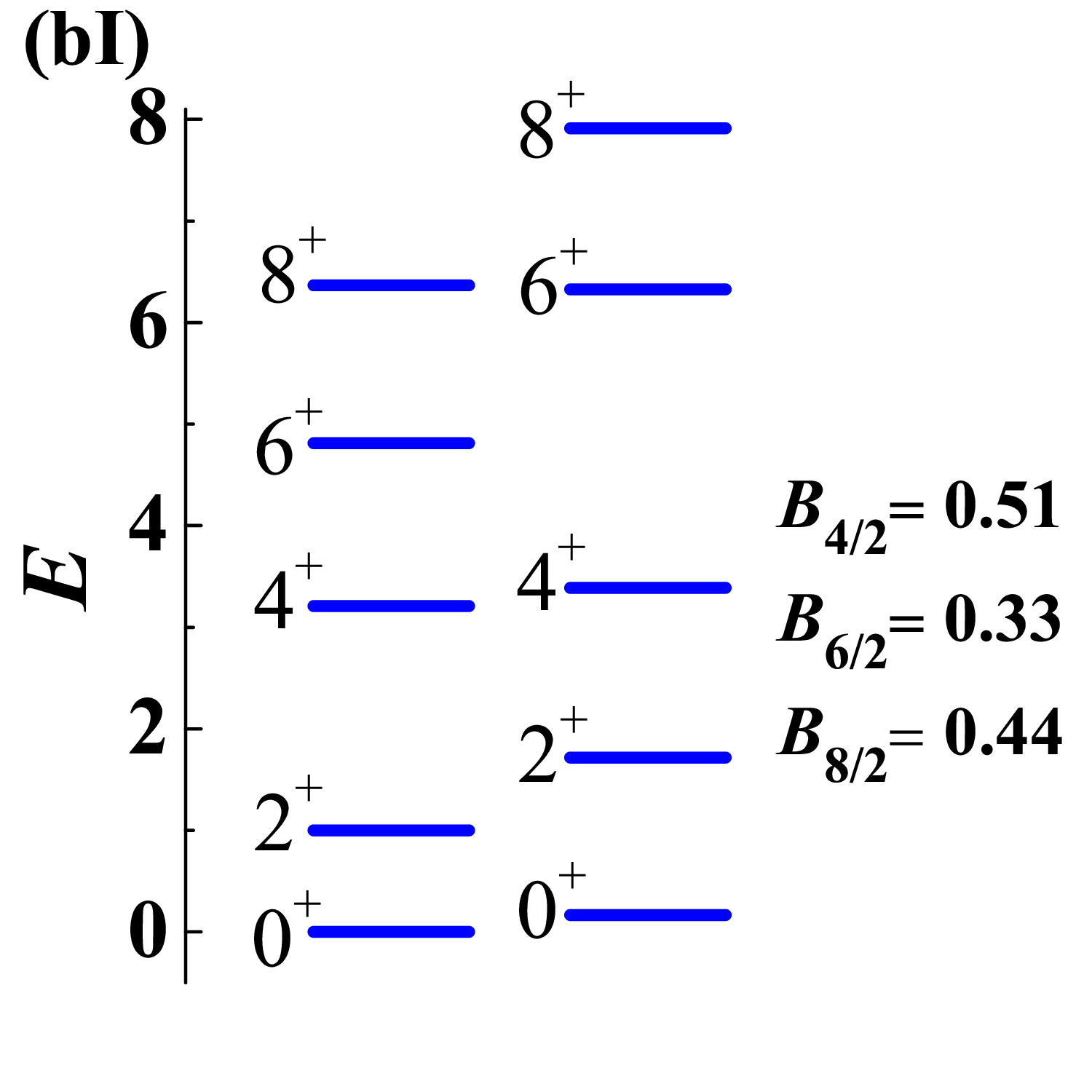}
\includegraphics[scale=0.16]{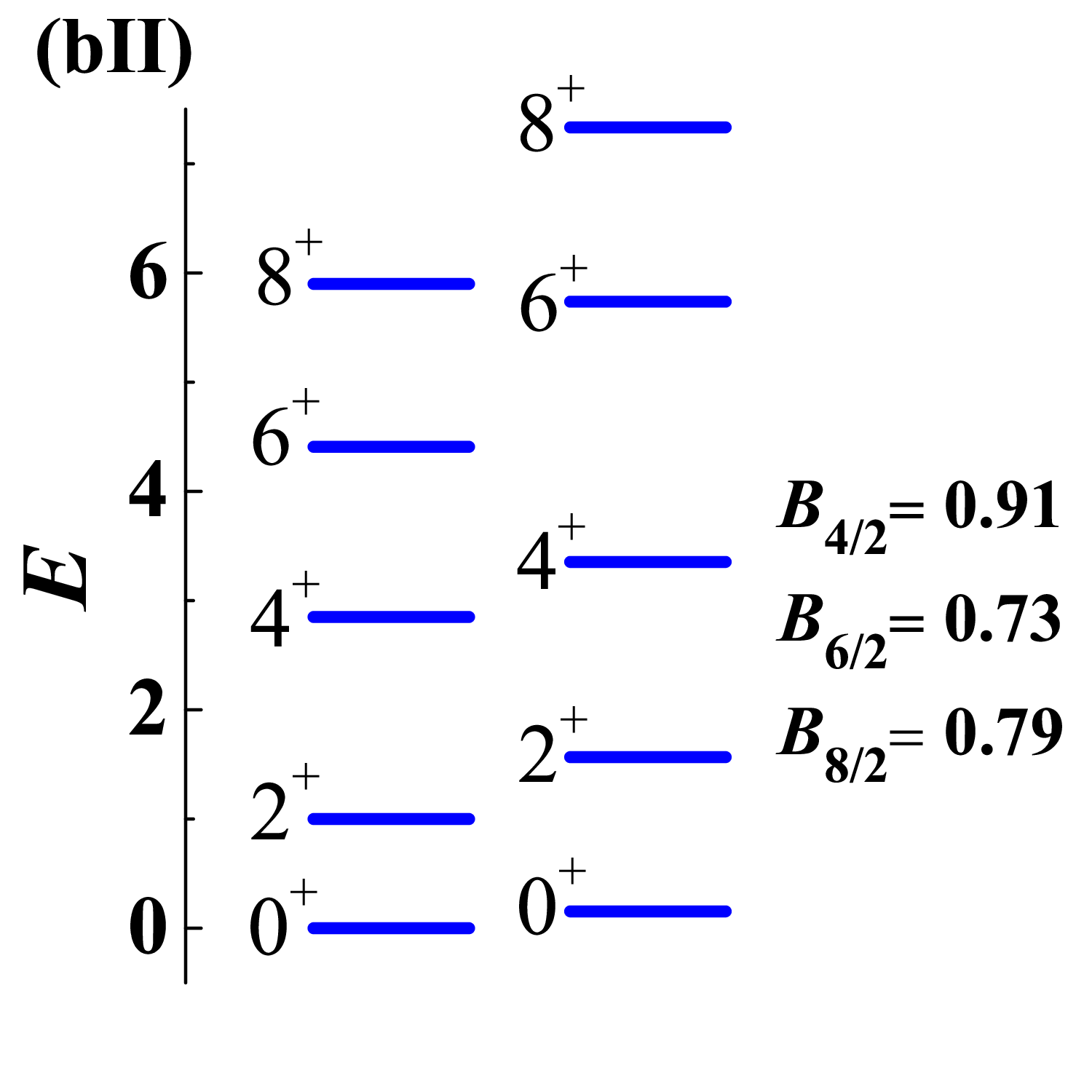}
\includegraphics[scale=0.16]{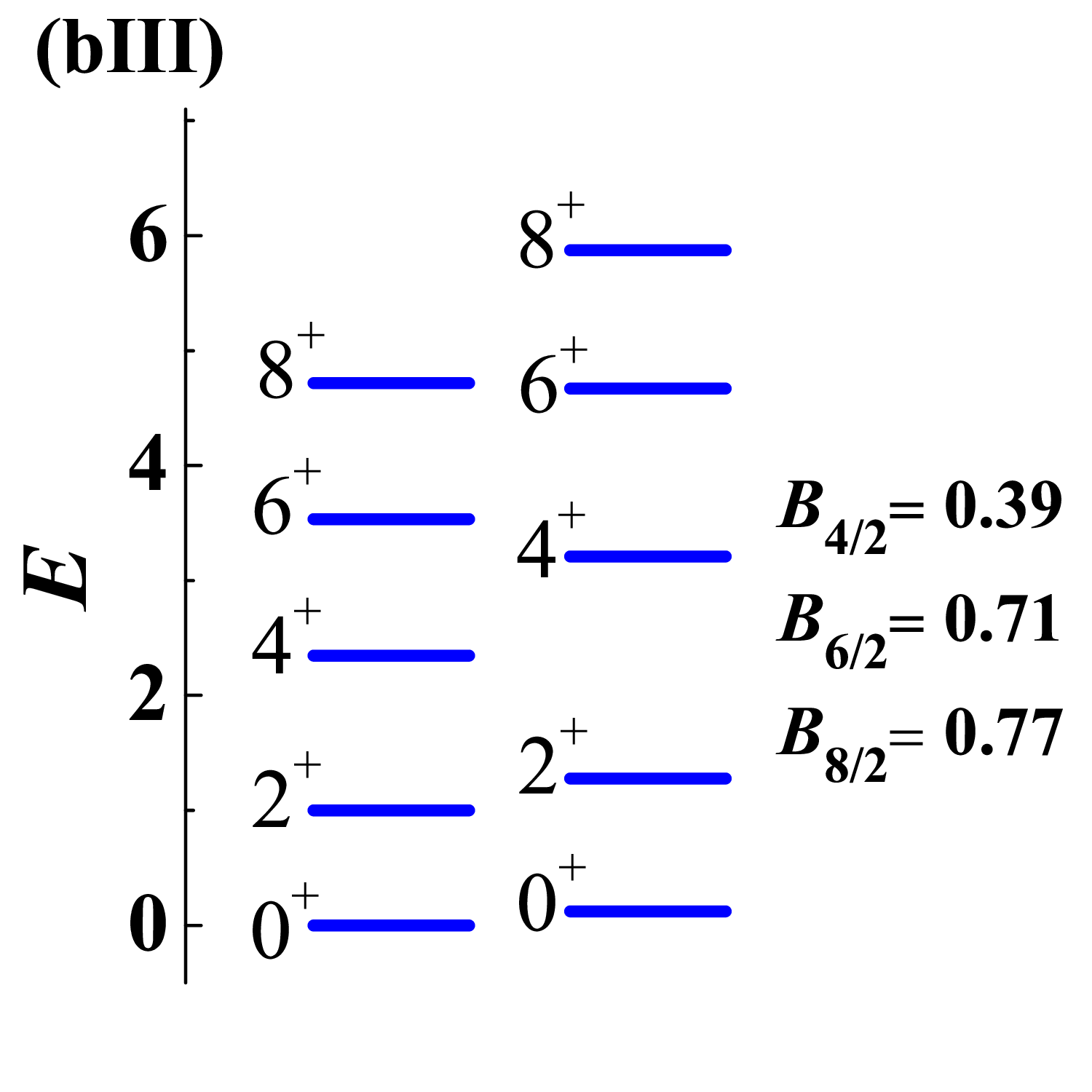}
\includegraphics[scale=0.16]{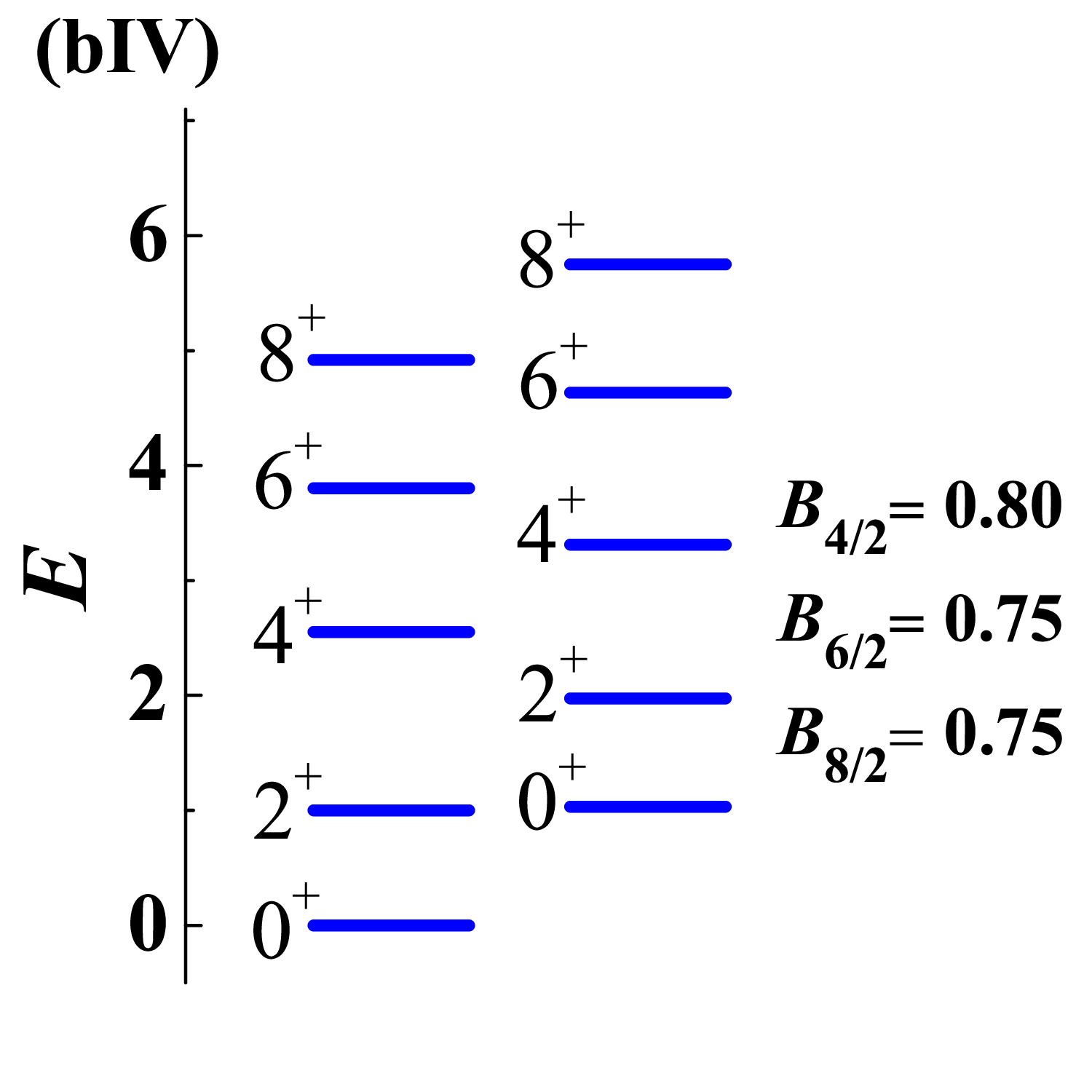}
\caption{Configuration mixing between U(5) and SU(3) with and without the inclusion of the $LQL$ term: (aI) The potential surface derived from (\ref{VMatrix}) in the weak mixing case. (aII) The same as in (aI), but for the intermediate mixing case. (aIII) The same in (aII), but obtained with the addition of the $LQL$ term. (aIV) The same as in (aIII), but for the strong mixing case. (bI) The level pattern (normalized to $E(2_1^+)=1.0$) corresponding to the case in (aI). (bII) The level pattern corresponding to the case in (aII). (bIII) The level pattern corresponding to the case in (aIII). (bIV) The level pattern corresponding to the case in (aIV). The adopted parameters in each panel are detailed in the text.}\label{F7}
\end{center}
\end{figure*}

To examine the impact of involving additional configurations we provide an example of the IBM-CM calculation incorporating three configurations by additionally introducing an intruder SU(3) configuration (4p-4h excitation) based on the mixing between the U(5) and O(6) modes shown in Fig.~\ref{F5}. In the calculations, the three involved configurations, U(5), O(6) and SU(3), are denoted as $A$, $B$ and $C$, respectively, with the nonzero parameters (in arbitrary units) set as $(a_1^A,~a_2^B,~a_2^C)=(1.0,-0.133,~-0.075)$, $(\Delta_B,~\Delta_C)=(19.28,~26.12)$, and $(\chi^A,~\chi^B,~\chi^C)=(0,~0,~-\sqrt{7}/2)$. Similarly, both the weak mixing case described by $\omega^{AB}=\omega^{BC}=0.001$ and strong mixing case described by $\omega^{AB}=\omega^{BC}=0.3$ are considered for comparison. The calculated potential surfaces and level patterns are shown in Fig.~\ref{F6}. As expected, the results demonstrate that incorporating more configurations leads to $B_{L/2}<1.0$ for more $L$ values, accompanied by an increase in the number of degenerated levels, as illustrated in Fig.~\ref{F6}(b1). Consistently, the $B(E2)$ anomaly features caused by level crossing in the weak mixing case disappear in the strong mixing case, as observed in Fig.~\ref{F6}(b2). Nonetheless, a common feature is that both cases yield numerous non-yrast states at low energy, resembling the band-mixing picture generated by the Hamiltonian involving the rotor-like terms adopted in prior~\cite{Zhang2022,Wang2020,Zhang2024,Pan2024,Teng2025,Zhang2025}.

Finally, we present an example of the combination of the configuration mixing with rotor-like terms based on the cases shown in Fig.~\ref{F3}, where configurations A and B are assumed to originate from the U(5) and SU(3) limits, respectively. For comparative purposes, Fig.~\ref{F7} displays the potential surfaces and level patterns obtained from four different cases to illustrate the influence of the $LQL$ term in the current configuration-mixing scheme. The parameters used are identical to those in Fig.~\ref{F3}, except for the additional nonzero parameters: $(b_2^B,\Delta,\omega)=(0.058,30.5,0.05)$ for case I, $(b_2^B,\Delta,\omega)=(0.065,30.71,0.15)$ for case II, and $(b_1^B,b_2^B,\Delta,\omega)=(-0.013,0.065,30.71,0.15)$ for case III, and $(b_1^B,b_2^B,\Delta,\omega)=(-0.08,0.058,30.3,0.7)$ for case IV. Compared to the cases present in Fig.~\ref{F3}, a notable observation in the present cases is the use of smaller off-set energies $\Delta$, which may lead to a relatively lower intruder configuration.

As seen from Fig.~\ref{F7}(bI), the current case exhibits more $B(E2)$ anomaly results compared to those observed in Fig.~\ref{F3}(b1). The mechanism for $B_{L/2}<1.0$ can be clearly explained by examining the $B(E2)$ characteristics in the two symmetry limit. The values of $B(E2;L_1^+\rightarrow (L-2)_1^+)$ in the SU(3) limit are typically much larger than those in the U(5) limit, as the leading order terms of these transitions are proportional to $N^2$ in the former and only to $N$ in the latter. For instances, at $N=10$, the SU(3) limit yields $B(E2;L_1^+\rightarrow (L-2)_1^+)/e^2=46,~64.29,~67.97,~66.87$ for $L=2,~4,~6,~8$, whereas the corresponding values in the U(5) limit are $10,~18,~24,28$ in the U(5) limit. If configuration mixing results in the $0_1^+$ and $2_1^+$ states belonging to the SU(3) limit-thus exhibiting a relative large $B(E2;2_1^+\rightarrow 0_1^+)$-while higher yrast states with $L\geq4$ belong to the U(5) limit, then in the zero mixing ($\omega=0$) limit, one obtains $B_{4/2}=0$ due to the prohibition of the $B(E2)$ transition between two configurations, and $0<B_{L/2}<1.0$ for all the ratios with $L>4$, due to the relatively large denominator and small numerators. This behavior is consistent with what is observed in Fig.~\ref{F7}(bI), where a small mixing parameter $\omega=0.05$ slightly increase $B_{4/2}$ above zero without significantly altering the other $B_{L/2}$ values. Importantly, the current mechanism responsible for $B_{L/2}<1$ differs from that associated with level crossing and does not result in energy degeneracies. Furthermore, compared to the parameter-sensitive outcomes caused by level crossing as shown in Fig.~\ref{F3}(b1), the present results for $B_{4/2}<1.0$ exhibit greater robustness against variations in model parameters.
When increasing $\omega$, as shown in Fig.~\ref{F7}(bII), all $B_{L/2}$ values rise, particularly for $B_{4/2}$. Another noticeable feature is that the energy of the $0_2^+$ state becomes significantly lower in these cases. By introducing the $LQL$ term into the Hamiltonian, as illustrated in Fig.~\ref{F7}(bIII), the results show that $B_{4/2}$ can be reduced to desired levels; however, this reduction does not alleviate the low $E(0_2^+)$. This limitation can be partially mitigated by simultaneously increasing $\omega$ and $b_1^B$, as observed from Fig.~\ref{F7}(bIV). The results clearly indicate that in this strong mixing regime, the level pattern associated with the $B(E2)$ anomaly becomes more regular and meanwhile does not exhibit any level degeneracies. In fact, the case presented in Fig.~\ref{F7}(bIV) is similar to the earlier mentioned example used to illustrate how rotor-like terms under strong configuration mixing can produce more instances of $B_{L/2}<1.0$. Additionally, as seen from Fig.~\ref{F7}, the potential surfaces across all four cases are very similar, and it is difficult to identify at the mean-field level any noticeable features associated with the $B(E2)$ anomaly in the present cases. Overall, the effects of configuration mixing on the $B(E2)$ structure can also be described as band mixing. Therefore, upon the band mixing associated with dynamical triaxility, configuration mixing appears to provide additional opportunities to generate $B(E2)$ anomaly in theory.

\begin{center}
\vskip.2cm\textbf{IV. Summary}
\end{center}\vskip.2cm

In summary, a theoretical analysis of the band-mixing mechanism for $B(E2)$ anomaly related to dynamical triaxiality and configuration mixing has been conducted within the framework of the IBM.
The results reveal that introducing the rotor-like term $LQL$ in the Hamiltonian may induce an evident change in the effective $\gamma$ deformation as angular momentum increase, thereby clarifying
the potential correlation between $B(E2)$ anomaly and triaxiality. Furthermore, the IBM-CM calculations suggest that mixing between normal and intruder bands can, to some extent, enhance the likelihood of the occurrence of $B(E2)$ anomaly. Although this result is obtained in a parameter-dependent manner, the currently available data cannot rule out this possibility for light Pt nucleus associated with $B(E2)$ anomaly, particularly given that configuration mixing appears to be essential for accurately describing Pt isotopes, as indicated by the prior IBM-CM studies~\cite{Dracoulis1994,King1998,Harder1997,Morales2008,Ramos2009,Ramos2011,Ramos2014}.

The present analysis provides necessary supplements to the previous IBM explanation of $B(E2)$ anomaly based on the band-mixing mechanism~\cite{Zhang2014,Zhang2022,Wang2020,Zhang2024,Pan2024,Teng2025,Zhang2025}, offering additional references for future experimental investigations of relevant neutron-deficient nuclei. Mostly recently, it has been demonstrated~\cite{Zhang2025II,Teng2025II} that triaxiality-related $B(E2)$ anomaly features can also be generated within the microscopic version of the IBM (IBM-2)~\cite{Iachellobook}, which distinguishes between protons and neutrons. It would be interesting to explore whether the conclusions drawn here can be extended to the IBM-2. Additionally, introducing $g$ bosons represents another way to produce triaxial deformation in the IBM~\cite{Kuyucak1991}, thus potentially providing an alternative pathway for generating $B(E2)$ anomaly. Work in this direction is in progress.
\bigskip

\begin{acknowledgments}
We wish to thank F. Pan and C. Qi for their stimulating discussions on the related topics, and Z. P. Li for his guidance on the mean-field calculation
using the covariant density functional theory with the PC-PK1 interactions. Support from the National Natural Science Foundation of China under No.12375113 is acknowledged.
\end{acknowledgments}




\begin{thebibliography}{99}


\bibitem{Bohrbook}A. Bohr and B. R. Mottelson, {\it Nuclear Structure II} (Benjamin, New York, 1975).


\bibitem{Grahn2016}T. Grahn {\it et al.}, Phys. Rev. C {\bf 94}, 044327 (2016).

\bibitem{Saygi2017}B. Say{\v g}{\i} {\it et al.}, Phys. Rev. C {\bf 96}, 021301(R) (2017).

\bibitem{Cederwall2018}B. Cederwall {\it et al.}, Phys. Rev. Lett. {\bf 121}, 022502 (2018).

\bibitem{Goasduff2019}A. Goasduff {\it et al.}, Phys. Rev. C {\bf 100}, 034302 (2019).

\bibitem{Zhang2021}W. Zhang {\it et al.}, Phys. Lett. B {\bf 820}, 136527 (2021).

\bibitem{Zanon2025}I. Zanon {\it et al.}, Phys. Rev. C {\bf 111}, 034323 (2025).

\bibitem{Zhang2022} Y. Zhang, Y. W. He, D. Karlsson, C. Qi, F. Pan, and J. P. Draayer, Phys. Lett. B {\bf  834}, 137443 (2022).

\bibitem{Wang2020}T. Wang, Europhys. Lett. {\bf 129}, 52001 (2020).

\bibitem{Zhang2024} Y. Zhang, S. N. Wang, F. Pan, C. Qi, and J. P. Draayer, Phys. Rev. C {\bf 110}, 024303 (2024).

\bibitem{Pan2024} F. Pan, Y. Zhang, Y. X. Wu, L. R. Dai, and J. P. Draayer, Phys. Rev. C {\bf 110}, 054324 (2024).

\bibitem{Teng2025} W. Teng, Y. Zhang, and C. Qi, Chin. Phys. C {\bf 49}, 014102 (2025).

\bibitem{Zhang2025} Y. Zhang and W. Teng, Phys. Rev. C {\bf 111}, 014324 (2025).

\bibitem{Iachellobook}
        F. Iachello and A. Arima, {\it The Interacting Boson Model}
        (Cambridge University, Cambridge, England, 1987).



\bibitem{Draayer1983}J. P. Draayer and K. J. Weeks, Phys. Rev. Lett. {\bf51}, 1422 (1983).

\bibitem{Draayer1989}J. P. Draayer, S. C. Park, and O. Casta{\~n}os, Phys. Rev. Lett. {\bf62}, 20 (1989).

\bibitem{Leschber1987}Y. Leschber and J. P. Draayer, Phys. Lett. B {\bf190}, 1 (1987).

\bibitem{Castanos1988}O. Casta{\~n}os, J. P. Draayer, and Y. Leschber, Z. Phys. A {\bf329}, 33
(1988).


\bibitem{Heyde1984}K. Heyde, P. Van Isacker, M. Waroquier, and J. Moreau, Phys. Rev. C {\bf29}, 1420 (1984).

\bibitem{Berghe1985}G. Vanden Berghe, H. E. De Meyer, and P. Van Isacker, Phys. Rev. C {\bf 32}, 1049 (1985).


\bibitem{Vanthournout1988}J. Vanthournout, H. De Meyer, and G. Vanden Berghe, Phys. Rev. C {\bf 38}, 414 (1988).

\bibitem{Vanthournout1990}J. Vanthournout, Phys. Rev. C {\bf 41}, 2380 (1990).

\bibitem{Smirnov2000} Y. F. Smirnov, N. A. Smirnova, and P. Van Isacker, Phys. Rev.
C {\bf61}, 041302(R) (2000).

\bibitem{Zhang2014} Y. Zhang, F. Pan, L. R. Dai, and J. P. Draayer, Phys. Rev. C {\bf  90}, 044310 (2014).

\bibitem{Teng2024}W. Teng, S. N. Wang, Y. Zhang, and L. Fortunato, Phys. Scr. {\bf  99}, 015305 (2024).

\bibitem{Wood2004}J. L. Wood, A-M. Oros-Peusquens, R. Zaballa, J. M. Allmond, and W. D. Kulp, Phys. Rev. C 70, 024308 (2004).

\bibitem{Allmond2008} J. M. Allmond, R. Zaballa, A. M. Oros-Peusquens, W. D. Kulp, J. L. Wood, Phys. Rev. C {\bf 78}, 014302 (2008).


\bibitem{Guzman2010}R. Rodr{\'i}guez-Guzm{\'a}n, P. Sarriguren, L. M. Robledo, and J. E. Garc{\'i}a-Ramos, Phys. Rev. C {\bf 81}, 024310 (2010).

\bibitem{Dracoulis1994} G. D. Dracoulis, Phys. Rev. C {\bf 49}, 3324 (1994).

\bibitem{King1998} S. L. King, {\it et al.}, Phys. Lett. B {\bf 443}, 82 (1998).

\bibitem{Harder1997} M. K. Harder, K. T. Tang, and P. Van Isacker, Phys. Lett. B {\bf 405}, 25 (1997).

\bibitem{Morales2008} I. O. Morales, A. Frank, C. E. Vargas, and P. Van Isacker, Phys. Rev. C {\bf 78}, 024303 (2008).

\bibitem{Ramos2009} J. E. Garc\'{i}a-Ramos and K. Heyde, Nucl. Phys. A {\bf 825}, 39 (2009).

\bibitem{Ramos2011} J. E. Garc\'{i}a-Ramos, V. Hellemans, and K. Heyde, Phys. Rev. C {\bf 84}, 014331 (2011).

\bibitem{Ramos2014} J. E. Garc\'{i}a-Ramos, K. Heyde, L. M. Robledo, and R. Rodr{\'i}guez-Guzm{\'a}n, Phys. Rev. C {\bf 89}, 034313 (2014).

\bibitem{Duval1982} P. D. Duval and B. R. Barrett, Nucl. Phys. A {\bf 376}, 213 (1982).


\bibitem{Castanos1984} O. Casta{\~n}os, A Frank, and P. Van Isacker, Phys. Rev. Lett. {\bf52}, 263 (1984).

\bibitem{Casten1984} R. F. Casten, A. Aprahamian, and D. D. Warner, Phys. Rev. Lett. {\bf29}, 356 (1984).

\bibitem{Elliott1986} J. P. Elliott, J. A. Evans, and P. Van Isacker, Phys. Rev. Lett. {\bf57}, 1124 (1986).

\bibitem{Vogel1996} O. Vogel, P. Van Isacker, A. Gelberg, P. von Brentano, and A. Dewald, Phys. Rev. C {\bf53}, 1660 (1996).


\bibitem{IC1981} P. Van Isacker and J. Q. Chen, Phys. Rev. C {\bf24}, 684 (1981).


\bibitem{Warner1983} D. D. Warner and R. F. Casten, Phys. Rev. C {\bf28}, 1798 (1993).

\bibitem{Sorgunlu2008} B. Sorgunlu and P. Van Isacker, Nucl. Phys. A {\bf  808}, 27 (2008).

\bibitem{Casten1985} R. F. Casten, P. Von Brentano, K. Heyde, P. Van Isacker, and J. Jolie, Nucl. Phys. A {\bf  439}, 289 (1985).

\bibitem{Ramos2000I} J. E. Garc\'{i}a-Ramos, C. E. Alonso, J. M. Arias, and P. Van Isacker, Phys. Rev. C {\bf  61}, 047305 (2000).

\bibitem{Ramos2000II} J. E. Garc\'{i}a-Ramos, J. M. Arias, and P. Van Isacker, Phys. Rev. C {\bf  62}, 064309 (2000).

\bibitem{Nomura2017I} K. Nomura, R. Rodr{\'i}guez-Guzm{\'a}n, Y. M. Humadi, L. M. Robledo, and H. Abusara, Phys. Rev. C {\bf 96}, 034310 (2017).

\bibitem{Nomura2017II} K. Nomura, R. Rodr{\'i}guez-Guzm{\'a}n, and L. M. Robledo, Phys. Rev. C {\bf 95}, 064310 (2017).

\bibitem{Davydov1958} A. S. Davydov and G. F. Filippov, Nucl. Phys., {\bf 8} 237 (1958).

\bibitem{Pan2003} F. Pan, J. P. Draayer, and Y. A. Luo, Phys. Lett. B, {\bf 576} 297 (2003).

\bibitem{Zhao2010} P. W. Zhao, Z. P. Li, J. M. Yao, and J. Meng, Phys. Rev. C, {\bf 82} 054319 (2010).

\bibitem{Dobes1985} J. Dobe\v{s}, Phys. Lett. B, {\bf 158} 97 (1985).

\bibitem{Werner2005} V. Werner, C. Scholl, and P. von Brentano, Phys. Rev. C, {\bf 71} 054314 (2005).


\bibitem{Fortunato2011} L. Fortunato, C. E. Alonso, J. M. Arias, J. E. Garc{\'i}a-Ramos, and A.
Vitturi, Phys. Rev. C {\bf 84}, 014326 (2011).


\bibitem{Heyde2011} K. Heyde and J. L. Wood, Rev. Mod. Phys. {\bf 83} 1467 (2011).


\bibitem{Gavrielov2019} N. Gavrielov, A. Leviatan, and F. Iachello, Phys. Rev. C, {\bf 99} 064324 (2019).


\bibitem{Frank2004} A. Frank, P. Van Isacker, and C. E. Vargas, Phys. Rev. C, {\bf 69} 034323 (2004).


\bibitem{Frank2006} A. Frank, P. Van Isacker, and F. Iachello, Phys. Rev. C, {\bf 73} 061302(R) (2006).

\bibitem{Zhang2025II}W. Teng, Y. Zhang, S. N. Wang, F. Pan, C. Qi, and J. P. Draayer,  Phys. Lett. B {\bf 865}, 139487 (2025).

\bibitem{Teng2025II} W. Teng, S. N. Wang, Y. Zhang, X. Z. Zhao, X. Deng, and X. T. Li, Chin. Phys. C {\bf 49}, 084106 (2025).


\bibitem{Kuyucak1991}S. Kuyucak and I. Morrison,  Phys. Lett. B {\bf 255}, 305 (1991).



\end{thebibliography}
\end{document}